\documentclass[11pt]{article}
\usepackage{eurosym}
\usepackage{caption}
\usepackage{diagbox}
\usepackage{bookmark}
\usepackage{authblk}
\usepackage{latexsym,amsmath,url,epsfig}
\usepackage{textcomp}
\usepackage{amsfonts,euscript}
\usepackage{amsmath}
\usepackage{amssymb}
\usepackage{amsfonts}
\usepackage{amsthm}
\usepackage{mathrsfs}
\usepackage[all]{xy}
\usepackage{epstopdf}
\usepackage{subfigure}
\usepackage{rotating}
\usepackage[title]{appendix}
\usepackage{multirow,rotating}
\usepackage{url}
\usepackage{algorithm}
\usepackage{algorithmicx}
\usepackage{algpseudocode}
\usepackage{subfigure}
\usepackage{setspace}
\usepackage{listings}
\usepackage{color}
\usepackage{tikz}
\usepackage{graphics}
\usepackage[english]{babel}
\usepackage{datetime}
\usepackage{subfigure}
\usepackage{booktabs}
\usepackage{rotfloat}
\usepackage[authoryear]{natbib}
\usepackage{framed}
\usepackage{natbib}
\usepackage{graphicx}
\usepackage{todonotes}

\bookmarksetup{
  numbered,
  addtohook={    \ifnum\bookmarkget{level}>1       \bookmarksetup{numbered=false}    \fi
  },
}
\usetikzlibrary{chains,shapes,decorations,arrows,calc,arrows.meta,fit,positioning}
\tikzset{three sided/.style={
        draw=none,
        append after command={
            [shorten <= -0.5\pgflinewidth]
            ([shift={(-0.5\pgflinewidth,-0.5\pgflinewidth)}]\tikzlastnode.north east)
        edge([shift={( 0.5\pgflinewidth,-0.5\pgflinewidth)}]\tikzlastnode.north west) 
              ([shift={( 0.5\pgflinewidth,-0.5\pgflinewidth)}]\tikzlastnode.north east)
        edge([shift={( 0.5\pgflinewidth,-.95\pgflinewidth)}]\tikzlastnode.south east)  
        
            ([shift={( -0.5\pgflinewidth,-0.5\pgflinewidth)}]\tikzlastnode.south west)
        edge([shift={(0.5\pgflinewidth,-0.5\pgflinewidth)}]\tikzlastnode.south east)
        }
    }
}
\tikzset{three sided2/.style={
        draw=none,
        append after command={
            [shorten <= -0.5\pgflinewidth]
          ([shift={( -0.5\pgflinewidth,-0.5\pgflinewidth)}]\tikzlastnode.north west)
        edge([shift={( -0.5\pgflinewidth,-.95\pgflinewidth)}]\tikzlastnode.south west)  
        
              ([shift={( 0.5\pgflinewidth,-0.5\pgflinewidth)}]\tikzlastnode.north east)
        edge([shift={( 0.5\pgflinewidth,-.95\pgflinewidth)}]\tikzlastnode.south east)  
        
            ([shift={( -0.5\pgflinewidth,-0.5\pgflinewidth)}]\tikzlastnode.south west)
        edge([shift={(0.5\pgflinewidth,-0.5\pgflinewidth)}]\tikzlastnode.south east)
        }
    }
}
\tikzset{three sided3/.style={
        draw=none,
        append after command={
            [shorten <= -0.5\pgflinewidth]
  ([shift={(-0.5\pgflinewidth,-0.5\pgflinewidth)}]\tikzlastnode.north east)
        edge([shift={( 0.5\pgflinewidth,-0.5\pgflinewidth)}]\tikzlastnode.north west) 
          ([shift={( -0.5\pgflinewidth,-0.5\pgflinewidth)}]\tikzlastnode.north west)
        edge([shift={( -0.5\pgflinewidth,-.95\pgflinewidth)}]\tikzlastnode.south west)  
        
              ([shift={( 0.5\pgflinewidth,-0.5\pgflinewidth)}]\tikzlastnode.north east)
        edge([shift={( 0.5\pgflinewidth,-.95\pgflinewidth)}]\tikzlastnode.south east)

        }
    }
}
\theoremstyle{definition}

\algdef{SE}[DOWHILE]{Do}{doWhile}{\algorithmicdo}[1]{\algorithmicwhile\ #1}

\allowdisplaybreaks

\begin{document}

\title{Singular Control of (Reflected) Brownian Motion: A Computational Method Suitable for Queueing Applications}
\author[1]{Baris Ata\thanks{baris.ata@chicagobooth.edu}}
\author[2]{J. Michael Harrison\thanks{mike.harrison@stanford.edu}}
\author[1]{Nian Si\thanks{niansi@chicagobooth.edu}}
\affil[1]{Booth School of Business, University of Chicago}
\affil[2]{Graduate School of Business, Stanford University}
%\affil[3]{IEDA, Hong Kong University of Science and Technology}
\date{ }
\maketitle

\begin{abstract}
Motivated by applications in queueing theory, we consider a class of
singular stochastic control problems whose state space is the $d$-dimensional positive orthant. The original problem is approximated by a
drift control problem, to which we apply a recently developed computational
method that is feasible for dimensions up to $d=30$ or more. To show that
nearly optimal solutions are obtainable using this method, we present
computational results for a variety of examples, including queueing network
examples that have appeared previously in the literature.
\end{abstract}

\section{Introduction}
\label{sec:intro}

In a recent paper \citep{ata2023drift} we described a new computational
method for drift control of reflected Brownian motion, illustrating its use
through applications to various test problems. Here we consider a class of
singular stochastic control problems (see below for a discussion of this
term) that are motivated by applications in queueing theory. Our approach is
to approximate a problem in that class by a drift control problem of the
type considered earlier, and we conjecture that, as a certain upper bound
parameter becomes large, the solution obtained is nearly optimal for the
original singular control problem. Numerical results will be presented here
that give credence to that conjecture, but we do not attempt a rigorous
proof.

\textbf{Singular Stochastic Control.} The modern theory of singular
stochastic control was initiated by \citet{benevs1980some}, whose work
inspired subsequent research by \citet{karatzas1983class}, %
\citet{harrison1983instantaneous} and others in the 1980s, all of which
focused on specially structured problem classes. In fact, no truly general
definition of singular control has been advanced to this day, but the term
is used in reference to problems having the three features specified in the
following paragraph, where we denote by $W(t)$ the state of the system at
time $t$.

(a) In the absence of control, $W = \{W(t), t \geq0\}$ evolves as a
diffusion process whose state space $S$ is a subset of $d$-dimensional
Euclidean space. In all the work of which we are aware, the drift vector and
covariance matrix of $W$ are actually assumed to be constant (that is, to
not depend on the current state), so the associated theory can be accurately
described as singular control of Brownian motion, or of geometric Brownian
motion in some financial applications, or of reflected Brownian motion in some queueing applications. (b) The system manager can enforce
instantaneous displacements in the state process $W$, possibly including
enforcement of jumps, in any of several fixed directions of control,
subject to the requirement that $W$ remain within the state space $S$. (c)
The controller continuously incurs running costs, usually called holding
costs in queueing theory applications, that are specified by a given
function $h(W(t))$. He or she also incurs costs of control that are
proportional to the sizes of the displacements enforced, with a different
constant of proportionality for each direction of control.

A control policy takes the form of a $p$-dimensional cumulative control
process $U = \{U(t), t \geq 0\}$ whose components are non-decreasing.
Because the cumulative cost of control in direction $i$ is proportional to $%
U_i(\cdot)$ by assumption, an optimal policy typically creates one or more
endogenous reflecting barriers. For example, %
\citet{harrison1983instantaneous} considered a one-dimensional control
problem (that is, $d =1$) where $W$ behaves as a $(\mu,\sigma)$ Brownian
motion $X = \{X(t), t \geq 0\}$ in the absence of control, the holding cost
function $h(\cdot)$ is convex, and the controller can displace $W(\cdot)$
either to the right at a cost of $r > 0$ per unit of displacement, or else
to the left at a cost of $l > 0$ per unit of displacement.

Assuming the objective is to minimize expected discounted cost over an
infinite planning horizon, an optimal policy enforces a lower reflecting
barrier at some level $a$, and an upper reflecting barrier at some level $b
> a$. That is, the two cumulative control processes $U_1 (\cdot)$ and $U_2
(\cdot)$, which correspond to rightward displacement and leftward
displacement, respectively, increase in the minimal amounts necessary to
keep $W(\cdot) = X(\cdot) + U_1 (\cdot) - U_2 (\cdot)$ confined to the
interval $[a,b]$. Both $U_1 (\cdot)$ and $U_2 (\cdot)$ are almost surely
continuous, but because the Brownian motion $X$ has unbounded variation in
every interval of positive length, each of them increases only on a set of
time points having Lesbegue measure zero. In this sense, the optimal
controls are almost surely singular with respect to Lebesgue measure; it is
that feature of optimal controls, in a different but similar problem
context, that led \citet{benevs1980some} to coin the term ``singular
control.''

To the best of our knowledge, the only previous works providing
computational methods for singular control problems are those by %
\citet{kushner1991numerical} and by \citet{kumar2004numerical}. In Section %
\ref{sec:BIGSTEP example} of this paper we consider a two-dimensional
problem and compare our results with those obtained using the
Kushner-Martins method, finding that the two solutions are virtually
indistinguishable. However, the methods propounded by both Kushner-Martins
and Kumar-Muthuraman are limited to low-dimensional applications, because
they rely on grid-based computations. To be specific, %
\citet{kushner1991numerical} employ a Markov chain approximation, whereas %
\citet{kumar2004numerical} rely on the finite-element method for analysis of
partial differential equations. In contrast, our simulation-based
approximation scheme is suitable for high-dimensional problems, two of which
will be analyzed in Sections \ref{sec:known} and \ref{sec:many}.

\textbf{The rest of this paper.} Section \ref{sec:general} describes in general terms the two closely related classes of problems to be treated here, and Section \ref{sec:drift} explains our approach to such problems. Readers will see that
our method can easily be extended to address other similar problems, some of which will be discussed later in Section \ref{sec:conclude}. Still, the
problem types described in Section \ref{sec:general} includes many important
applications, and they are general enough to prove the viability of our
approach via drift control approximations. In Section \ref{sec:known} our
method is applied to a specially structured family of singular control
problems that can be solved analytically. The numerical results presented
there indicate that a high degree of accuracy is achievable with the method,
and that it is computationally feasible for problems of dimension 30 or more.

Sections \ref{sec:tandem} through \ref{sec:many} are devoted to stochastic
control problems associated with queueing network models, also called
stochastic processing networks, cf. \citet{dai2020processing}. More
specifically, the problems considered in those sections involve dynamic
control of networks in the ``heavy traffic'' parameter regime, for which
singular control approximations have been derived in previous research.
Analysis of such examples involves two steps: first, numerical solution of
the approximating singular control problem; and second, translation of the
singular control solution into an implementable policy for the original
discrete-flow model. The second of those steps will be left as an open
problem for one of our queueing network examples, because it involves what
is essentially a separate research domain. However, we intend to continue
analysis of that example in future work.

Section \ref{sec:conclude} contains concluding remarks on various topics, including the
potential for generalization beyond the problems described in Section \ref{sec:general}. Finally, there are several short appendices that provide added technical detail on subjects treated or referred to in the body of the paper.

The code used for the examples in this paper is available in \url{https://github.com/nian-si/SingularControl}. 
\section{Two classes of singular control problems to be treated} \label{sec:general} 

We consider a stochastic system with a $d$-dimensional 
\textit{state process} $W=\{W(t),t\geq 0\}$ and a $p$-dimensional \textit{%
control} $U=\{U(t),t\geq0\}$. Two distinct problem formulations are to be treated, each of them generated from the following primitive stochastic element:
\begin{align}
&X=\{X(t),t\geq 0\} \text{ is a $d$-dimensional Brownian motion with drift
vector $\xi$,} \notag \\
&\quad \text{non-singular covariance matrix A, and } X(0)=0.
\label{problem:state:X}   
\end{align}
\textbf{Formulation with exogenous reflection at the boundary.} In this problem $W$, $U$ and a $d$-dimensional \textit{boundary pushing process} $Y$ jointly satisfy the following relationships and restrictions: 
\begin{align}
&W(t)=w+X(t)+GU(t)+RY(t) \text{ for all } t\geq0,
\label{new:problem:state:w} \\
&U(\cdot) \text{ is adapted to } X, \text{ right-continuous and
non-decreasing with } U(0)\geq 0 \text{,}  \label{new:problem:U} \\
&W(t)\geq 0 \text{ for all } t \geq0,  \label{new:eq:positive_state} \\
&Y=\{Y(t),t \geq 0\} \text{ is adapted to } X, \text{ continuous and non-decreasing with } Y(0) = 0, \text{ and}
\label{new:boundary:Y1} \\
&Y_i(\cdot) \text{ only increases at times } t\geq 0 \text{ when } W_i(t)=0 
\text{ } (i=1,\ldots,d).  \label{new:boundary:Y2}
\end{align}
Here $w \geq 0$ is a given initial state, $G$ is a $d \times p$ matrix whose columns are interpreted as feasible \textit{directions of control}, and $R$ is a $d \times d$ \textit{reflection matrix} that has the form
\begin{equation}
R=I-Q \text{ where $Q$ is non-negative and has spectral radius $\rho(Q)<1$}. \label{assumption:M-matrix}
\end{equation}
Conditions (\ref{new:problem:state:w}) through (\ref{assumption:M-matrix}) together define $Y$ in terms of $X$ and $U$, specifying that $W$ is instantaneously reflected at the boundary of the orthant, and more specifically, that the direction of reflection from the boundary surface $W_i=0$ is the $i^{th}$ column of $R$. See below for further discussion.

Finally, we take as given a \textit{holding cost function} $h:\mathbb{R}^d_+ \rightarrow \mathbb{R}$, a vector $c\in \mathbb{R}_+^p$ of \textit{control
costs}, a vector $\pi \in \mathbb{R}_+^d$ of \textit{penalty rates} associated with pushing at the boundary, and an interest rate $\gamma > 0$ for
discounting. The system manager's objective is then to 
\begin{equation}
\text{minimize } \mathbb{E}\left \{\int_0^\infty e^{-\gamma t} \left[ h(W(t))dt + c\cdot
d U(t)+ \pi \cdot d Y(t) \right]\right\}.  \label{min:objective}
\end{equation}

The stochastic control problem (\ref{new:problem:state:w})-(\ref{min:objective}) is identical to the one studied in our earlier paper \citep{ata2023drift} except that there we imposed the following additional restriction: each component of the monotone control $U$ was required to be absolutely continuous with a density bounded above by a given constant $b>0$. It is the absence of that requirement (uniformly bounded rates of control), together with the linear cost of control assumed in (\ref{min:objective}), that distinguishes the current problem as one of singular control. It should be noted that the vector of penalty rates in (\ref{min:objective}) was denoted by $\kappa$ rather than $\pi$ in our earlier paper \citep{ata2023drift}, because $\pi$ was reserved there for a different use. Here that competing use does not exist, and we choose the notation $\pi$ because of its value as a mnemonic for "penalty." 

For each $j=1,\ldots,p$ and $t \geq 0$, one interprets $%
U_j (t)$ as the cumulative amount of displacement effected over the time
interval $[0, t]$ in the $j^{th}$ available direction of control (that is,
in the direction specified by the $j^{th}$ column of $G$), and one
interprets $c_j U_j (t)$ as the associated cumulative cost. Condition (\ref{new:problem:U}) allows instantaneous displacements in any of the feasible
directions of control, represented by jumps in the corresponding components
of $U$, including jumps in one or more of those directions at $t=0$,
represented by positive values for different components of $U(0)$.

Our use of the letter $W$ to denote system state, as opposed to the letter $%
Z $ that was used for that purpose in \citet{ata2023drift}, is motivated by
applications in queueing theory (see Sections \ref{sec:tandem} through \ref%
{sec:many}), where it is mnemonic for \textit{workload}. In general, those
queueing theoretic applications give rise to workload processes whose state
space is a $d$-dimensional polyhedral cone, as opposed to the orthant that
is specified by (\ref{new:eq:positive_state}) as the state space for our control
problem here. However, a polyhedral cone can always be transformed to an
orthant by a simple change of variables, so one may say that here we present
our control problem in a convenient \textit{canonical form}, with the goal
of simplifying the exposition.

A square matrix $R$ of the form (\ref{assumption:M-matrix}) is
called a \textit{Minkowski matrix} in linear algebra (or just \textit{%
M-matrix} for brevity). It is non-singular, and its inverse is given by the
Neumann expansion $R^{-1}=I+Q+Q^2+\cdots$. A process $U$ that satisfies conditions (\ref{new:problem:state:w}) through (\ref{new:boundary:Y2}) will be called a \textit{feasible control}, and by assuming that $R$ is an M-matrix we ensure the existence  of at least one feasible control, as follows. If one simply sets $U(\cdot)=0$, then a result by \citet{harrison1981reflected} shows that there exist $d$-dimensional processes $W$ and $Y$ that jointly satisfy (\ref{new:problem:state:w}) through (\ref{new:boundary:Y2}). Moreover, $W$ and $Y$ are unique in the pathwise sense, and $Y$ is adapted
to $X$. Thus the control $U(\cdot)=0$ is feasible. The resulting state process $W$ is called a \textit{reflected Brownian motion} (RBM) with reflection matrix $R$. 

\textbf{Formulation without exogenous reflection at the boundary.} In our second problem formulation, the assumption of exogenous reflection at the boundary is removed, so system dynamics are specified as follows:
\begin{align}
&W(t)=w+X(t)+GU(t) \text{ for all } t\geq0,  \label{problem:state:w} \\
&U(\cdot) \text{ is adapted to } X, \text{ right-continuous, and
non-decreasing with } U(0)\geq 0 \text{, and }  \label{problem:U} \\
&W(t)\geq 0 \text{ for all } t \geq0 .  \label{eq:positive_state}
\end{align}
As before, $G$ is a $d\times p$ matrix whose columns are interpreted as feasible 
directions of control, but now we assume that $p \geq d$ and 
\begin{equation}
\text{the first $d$ columns of $G$ form an M-matrix that we denote hereafter as $R$.}
\label{submatrix assumption}
\end{equation}
Finally, given a holding cost function $h(\cdot)$ and control cost vector $c$ as before, the system manager's objective is to
\begin{equation}
\text{minimize } \mathbb{E}\left \{\int_0^\infty e^{-\gamma t} \left[ h(W(t))dt + c\cdot
d U(t) \right]\right\}.  \label{min:objective:new}
\end{equation}

Assumption (\ref{submatrix assumption}) is satisfied in most queueing applications, as we shall see in Sections \ref{sec:tandem} through \ref{sec:many}, and it guarantees the existence of a feasible control in the following obvious manner. First, given the underlying Brownian motion $X$, let $W$ be the corresponding RBM and $Y$ its associated boundary pushing process, both constructed from $X$ as in \citet{harrison1981reflected} using $R$ as the reflection matrix. Then define a $p$-dimensional control $U$ by taking $U_i=Y_i$ for $%
i=1,\ldots,d$ and $U_i =0$ for $i=d+1,\ldots,p$.
This specific control $U$ is continuous
but in general not absolutely continuous. Of course, a feasible control that
minimizes the objective in (\ref{min:objective:new}) may be very different in character.

\section{Approximation by a drift control problem}
\label{sec:drift}
This section explains our computational approach to each of the problem classes introduced in Section \ref{sec:general}, but taking them in reverse order. Readers will see that all of our examples can be viewed as applications of the formulation \textit{without} exogenous reflection at the boundary, which is one reason for treating it first.

\textbf{Formulation without exogenous reflection at the boundary.} Let us first consider the singular control problem \textit{without} exogenous reflection at the boundary, defined by system relationships (\ref{problem:state:w})-(\ref{eq:positive_state}), our crucial assumption (\ref{submatrix assumption}) on the control matrix $G$, and the objective (\ref{min:objective:new}). To approximate this problem, we restrict attention to controls $U$ that satisfy conditions (\ref{def:U_1d})-(\ref{boundary:Y2again})
below, where $W$ is again defined in terms of $X$ and $U$ via (\ref{problem:state:w}), $W$ must again be confined to the non-negative orthant as in (\ref{eq:positive_state}), and the objective (\ref{min:objective:new}) remains unchanged. The first restriction on controls $U$ is that they have the following form: for each $t \geq 0$, 
\begin{align}
&U_i(t)=\int_0^t \theta_i(s) ds + Y_i(t) & \text{ for } i=1,\ldots,d \text{
and}  \label{def:U_1d} \\
&U_j(t)=\int_0^t\theta_j(s)ds & \text{ for } j=d+1,\ldots,p,
\label{def:U_d+1p}
\end{align}
where $\theta(\cdot)$ is a $p$-dimensional \textit{drift control}, and $%
Y(\cdot)$ is a corresponding $d$-dimensional \textit{boundary control} to be
specified shortly. Second, it is required that $\theta(\cdot)$ be adapted to 
$X$ and satisfy 
\begin{equation}
0\leq \theta_i(\cdot)\leq b \text{ for each } i =1,\ldots,p, \text{ where }
b>0 \text{ is a given constant.}  \label{eq:upperbound_b}
\end{equation}
Third, $Y(\cdot)$ is defined in terms of $X$ and $\theta$ by the following
relationships, which are identical to (\ref{new:boundary:Y1}) and (\ref{new:boundary:Y2}): 
\begin{align}
&Y(\cdot) \text{ is adapted to } X, \text{ continuous and non-decreasing with } Y(0)=0, \text{ and}
\label{boundary:Y1again} \\
&Y_i(\cdot) \text{ only increases at times } t\geq 0 \text{ when } W_i(t)=0 
\text{ } (i=1,\ldots,d).  \label{boundary:Y2again}
\end{align}
By combining (\ref{def:U_1d}) and (\ref{def:U_d+1p}) with our assumption (%
\ref{submatrix assumption}) about the form of the control matrix $G$, one
sees that the main system equation (\ref{problem:state:w}) can be rewritten
as follows when controls are of the restrictive form considered here: 
\begin{equation}
W(t)=w+X(t)+G\int_0^t \theta(s) ds + RY(t) \text{ for all } t \geq 0.
\label{eq:main}
\end{equation}%
As noted earlier, conditions (\ref{boundary:Y1again}) and (\ref%
{boundary:Y2again}) are the standard means of specifying instantaneous
reflection at the boundary of the orthant. More specifically, the direction
of reflection from the boundary surface $W_i=0$ is the $i^{th}$ column
of $R$, or equivalently, the $i^{th}$ column of $G$ $(i=1,\ldots,d)$.
One interprets $Y_i (t)$ as the cumulative amount of displacement effected
in that direction over the time interval $[0, t]$ in order to prevent $W$
from exiting the positive orthant. Note that the objective (\ref{min:objective:new}) associates a cost of $c_i$ with each unit of increase in $Y_i(\cdot)$, just as it continuously accumulates drift-related costs at rate $c_i\theta_i$ per time unit ($i=1, \ldots,d$). 

In our current setting, the boundary controls are redundant in an obvious
sense: the $i^{th}$ column of $R$ is identical to the $i^{th}$
column of $G$, so ``reflection'' at the boundary surface $W_i=0$ displaces
the system state $W$ in the same direction as does the $i^{th}$ component of
the system manager's drift control $\theta(\cdot)$, and we associate the
same linear cost rate with increases in $Y_i (\cdot)$ as we do with positive
values of $\theta_i (\cdot) $. The only reason for including reflection at
the boundary as a ``backup capability'' in our formulation is to ensure that
feasibility can be maintained despite the upper bound $b$ on allowable drift
rates. In the limit as $b\uparrow \infty$, one expects that backup
capability to be increasingly irrelevant. For example, if displacement by
means of $Y_i (\cdot)$ were made even slightly more expensive than
displacement by means of $\theta_i (\cdot)$, one would expect increases in $%
Y_i (\cdot)$ under an optimal policy to be rare and small as $b$ becomes
large.

The approximating drift control problem specified immediately above is of the form considered in our earlier study \citep{ata2023drift}, where a computational method for its solution was developed and illustrated. As in that study, attention will be restricted here to stationary Markov policies, by which we mean that 
\begin{equation}
\theta (t)=u(W(t)), \, t\geq 0,\text{ for some measurable policy function }u:%
\mathbb{R}_{+}^{d}\rightarrow \lbrack 0,b]^{p}.
\end{equation}
We offer a brief summary of the method described in \citet{ata2023drift} and detail the hyperparameters applied in Appendix \ref{app:our-policies}.

\textbf{Formulation with exogenous reflection at the boundary.} Secondarily, let us consider the problem \textit{with} exogenous reflection at the boundary, defined earlier by the system relationships (\ref{new:problem:state:w})-(\ref{new:boundary:Y2}), assumption (\ref{assumption:M-matrix}) on the form of the reflection matrix $R$, and the objective (\ref{min:objective}). In this case we restrict attention to controls $U$ of the form
\begin{equation}
U(t)=\int_0^t \theta(s) ds, \label{eq:def:U} \\
\end{equation}
where $\theta(\cdot)$ is a $p$-dimensional drift control adapted to 
$X$ and satisfying 
\begin{equation}
0\leq \theta_i(\cdot)\leq b \text{ for each } i =1,\ldots,p, \text{ where }
b>0 \text{ is a given constant.}  \label{eq:new_upperbound}
\end{equation}
Given a control $U$ of this form, we define a corresponding boundary pushing process $Y$ and state process $W$ via (\ref{boundary:Y1again})-(\ref{eq:main}) as before. Finally, the objective for our approximating drift control formulation is again given by (\ref{min:objective}), which involves a vector $\pi$ of penalty rates associated with the various components of $Y$, separate from the cost rate vector $c$ associated with the endogenous control $U$. 

As mentioned earlier, we conjecture that the solution obtained from these drift control problems is nearly optimal for the original singular control problem as $b \uparrow \infty$. Although we don't attempt a proof of this conjecture, similar results have been proved in the literature. For example, \cite{menaldi-taksar1989} demonstrated that one can approximate a certain multi-dimensional singular control problem with no state space constraints by a sequence of drift control problems; see also \cite{williams-chow-menaldi1994} for related results. More recently, \cite{zhong2024} provides an approximation result for a singular control problem in the orthant.

\section{Comparison with known solutions} \label{sec:known}

Let us consider a one-dimensional ($d=1$) singular control problem with $p=2$ directions of control (up and down, or right and left), initially framing the problem as one without exogenous reflection at the boundary. To be specific, let us assume the following problem data: 
\begin{align*}
 \xi=0, \, A=[1],\, G=[1,-1], \, c=(0,1), \, h(w)=2w \textrm{ and } \gamma = 0.1.    
\end{align*}
That is, we solve the following singular control problem: 
\begin{align*}
& \min_{U(\cdot)}\mathbb{E}\left\{ \int_{0}^{\infty }e^{-0.1 t}\left[ 2W(t)+dU_2(t)%
\right] \right\} \\
&\textrm{subject to}\\
&W(t) =w+X(t)+ U_1(t) - U_2(t), \  t\geq 0, \\
& X=\{X(t),t\geq 0\}\text{ is a one-dimensional standard Brownian motion,} 
\notag \\
& U(\cdot )\text{ is adapted to }X,\text{ right-continuous and
non-decreasing with }U(0)\geq 0\text{, and } \\
& W(t)\geq 0\text{ for all }t\geq 0.
\end{align*}%
Because the holding cost function $h(\cdot)$ is positive and increasing, it is obvious that control component $U_1$, which pushes the state process $W$ upward or rightward, should be minimal. Equivalently stated, $U_1$ should be structured to enforce a lower reflecting barrier at zero. Thus the problem under discussion can be reframed as one of singular control \textit{with} exogenous reflection at the boundary, and it is analyzed (for general problem data) in those terms in Appendix \ref{appendix:deri:1d}. There we show that the optimal choice for $U_2$ enforces an upper reflecting barrier at a threshold value $w^*>0$, so the optimal policy confines $W$ to a closed interval $[0,w^*]$, and we provide a formula for $w^*$.

\textbf{Comparing solutions with different values for the upper bound $b$.} Now consider the drift control approximation (with exogenous reflection at the boundary) discussed in Section \ref{sec:general}, specialized to the one-dimensional problem described above. The analytic solution of that problem was derived in Appendix D.2 of our previous paper \citep{ata2023drift}. The optimal policy is of a bang-bang type, featuring a specific threshold level: the optimal control employs a downward drift at the maximum allowable rate $b$ when $W(\cdot)$ is above the threshold, and employs zero drift otherwise. Table \ref{tab:analytic_b} compares the optimal threshold values, and the objective values $V(0)$ achieved, for the drift control approximations with various values of $b$ and for our "exact" singular control formulation. It is notable that even with the very tight bound $b=5$, the approximation error expressed in terms of the system manager's objective is less than 1\%.
\begin{table}[tbh]
\centering
\begin{tabular}{ccccccc}
\toprule $b$ & 2 & 5 & 10 & 20 & 100 & Singular \\ 
\midrule 
Threshold & 0.52 & 0.63 & 0.67 & 0.70 & 0.72 & 0.72 \\
 $V(0)$ & 14.73 & 14.09 & 14.00 & 13.97 & 13.96 & 13.96 \\ 
\bottomrule % &  &  &  &  &  & 
\end{tabular}%
\caption{Analytic solutions with different upper bounds $b$ of the
one-dimensional problem.}
\label{tab:analytic_b}
\end{table}

\begin{figure}[th]
\centering	
\begin{tikzpicture}[start chain=going right,>=latex,node distance=1pt]
		\node[three sided,minimum width=1.2cm,minimum height =0.8cm,on chain]  (wa2)  {$1$};
		
		\node[draw,circle,on chain,minimum size=1cm] (se2) {$\mu$};
		
		\draw[<-] (wa2.west) -- +(-20pt,0) node[left] {$\lambda$}; 
		\draw[->] (se2.east) -- node[above] {}  +(25pt,0);
		
		\node[three sided,minimum width=1.2cm,minimum height =0.8cm,on chain]  (wa22) [below = of wa2,yshift = -1cm] {$K$};
		
		\node[draw,circle,on chain,minimum size=1cm] (se22) {$\mu$};
		
		\draw[<-] (wa22.west) -- +(-20pt,0) node[left] {$\lambda$}; 
		\draw[->] (se22.east) -- node[above] {}  +(25pt,0);
		\path (wa2.south) -- node[auto=false,yshift = 0.1cm]{\vdots} (wa22.north);
	\end{tikzpicture}
\caption{A decomposable parallel-server queueing network.}
\label{fig:parallel}
\end{figure}
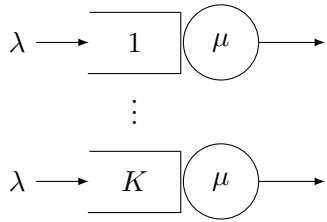

\textbf{A decomposable multi-dimensional singular control problem.} Let us now consider a $d$-dimensional problem of the form specified in Section \ref{sec:general} (\textit{without} exogenous reflection at the boundary) and the following data: control dimension $p=2d$, drift vector $\xi=0$, covariance matrix $A = I_{d\times d}$, control cost vector $c=(0,0,\ldots,0,1,1,\ldots,1)$, holding cost function $h(w)=2w_1+\cdots+2w_d$, interest rate $\gamma=0.1$, and control matrix 
\begin{equation*}
G=\left[ 
\begin{array}{cccccccc}
1 &  &  &  & -1 &  &  &  \\ 
& 1 &  &  &  & -1 &  &  \\ 
&  & \ddots &  &  &  & \ddots &  \\ 
&  &  & 1 &  &  &  & -1%
\end{array}%
\right]
\end{equation*}
This means that the $d$ components of $W$ are controlled independently, and the control problem for each individual component is the one-dimensional singular control problem discussed immediately above. This can be interpreted as a heavy-traffic approximation for a queueing network model where $d$ single-server queues operate independently and in parallel (see Figure \ref{fig:parallel}); a system manager can shut off the input to any one of the queues at any time, incurring a cost of 1 per unit of time that input is denied, and continuously incurs a cost of 2 per job held in the queue or in service.

We solved this 30-dimensional problem using our computational method with an upper bound value $b=10$, making no use of the problem's decomposability, and then we estimated the objective value $V(0)$ for our computed solution using Monte Carlo simulation. (That is, we repeatedly simulated the evolution of the Brownian system model with starting state $w=0$ and operating under our computed optimal policy, recording the average objective value and an associated confidence interval.) The result is reported in Table \ref{tab:parallelQ}, where we also record the simulated performance of the known analytic solution. Of course, the performance figures reported in Table \ref{tab:parallelQ} are subject to simulation and discretization errors.  The run-time for our method is about
twenty-four hours in this case using a 20-CPU core computer.
\begin{table}[!htb]
\centering
\begin{tabular}{cc}
\toprule Analytic optimal solution & Our method \\ 
\midrule 417.4 $\pm$ 0.2 & 416.8 $\pm$ 0.2 \\ 
\bottomrule% & 
\end{tabular}%
\caption{Simulation performance for the 30-dimensional parallel-server
network.}
\label{tab:parallelQ}
\end{table}

\section{Tandem queues example}

\label{sec:tandem} 
\begin{figure}[!htb]
\centering
\begin{tikzpicture}[start chain=going right,>=latex,node distance=1pt]
		% the rectangular shape with vertical lines
		\node[three sided,minimum width=1.5cm,minimum height = 1cm,on chain] (wa) {1};

		% the circle
		\node[draw,circle,on chain,minimum size=1cm] (se) {$S_1$};

		\node[three sided,minimum width=1.5cm,minimum height = 1cm,on chain]  (wa5)[right =of 
		se, xshift=1cm]  {2};
		\node[draw,circle,on chain,minimum size=1cm] (se2) {$S_2$};
		\node [above = of se] {$\mu$};
		\node [above = of se2] {$\mu$};

		% the arrows and labels
		\draw[->] (se) edge node[above] {} (wa5.west);
		\draw[->] (se2.east) -- +(30pt,0);
		
		\draw[<-] (wa.west) -- +(-20pt,0) node[left] {$\lambda_1$};

	\end{tikzpicture}
\caption{A network of tandem queues.}
\label{fig:tandem}
\end{figure}
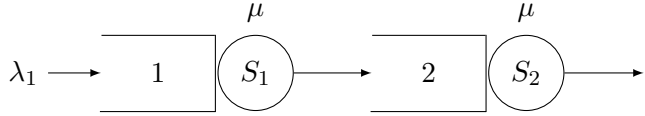
Figure \ref{fig:tandem} pictures a network model with two job classes and
two single-server stations. Class 1 jobs arrive from outside the system
according to a Poisson process with rate $\lambda$, and they are served at
station 1 on a first-in-first-out (FIFO) basis. After completing service at
station 1, they become class 2 jobs and proceed to station 2, where they are
again served on a FIFO basis. Class 2 jobs leave the system after completing
service. For jobs of each class there is a similarly numbered buffer
(represented by an open-ended rectangle in Figure \ref{fig:tandem}) where
jobs reside both before and during their service.

For simplicity, we assume that service times at each station are i.i.d.
exponential random variables with mean $m<\lambda^{-1}$, and furthermore, any service can be interrupted at any time and resumed later without any efficiency loss. We denote by $%
Q_k(t) $ the number of class $k$ jobs in the system at time $t$ $(k=1,2)$,
calling this a "queue length process," and when we speak of the "system
state" at time $t$, that means the pair $Q(t)=(Q_1(t),Q_2(t))$. There is a
holding cost rate of $h_k$ per class $k$ job per unit of time, so the
instantaneous holding cost rate at time $t$ is $h_1 Q_1(t)+h_2 Q_2(t)$.

We shall assume the specific numerical values $\lambda =0.95$ and $m=1$,
with holding costs $h_{1}=1$ and $h_{2}=2$, and with interest rate $r
=0.01$ for discounting. The system manager's control policy specifies
whether the two servers are to be working or idle in each system state. If
one is given holding costs such that $h_{1}\geq h_{2}>0$, then an optimal
policy is to have each server keep working so long as there are customers
present in that server's buffer, but with the data assumed here, it is more
expensive to hold class 2 than class 1, so there is a potential motivation
to idle server 1 even when there are customers present in its buffer. The
system manager seeks a control policy to minimize total expected discounted
holding costs over an infinite planning horizon.

Following the development in \citet{harrison1988brownian}, we imagine this
network model as part of a sequence indexed by $n=1,2,\cdots$ that
approaches a heavy traffic limit. In the current context, the assumption of
"heavy traffic" can be made precise as follows: there exists a large integer 
$n$ such that $\sqrt{n}(\lambda-m^{-1})$ is of moderate absolute value. With
the system parameters specified above, a plausible value of the associated
sequence index is $n=400$, which gives $\lvert \sqrt{n}(\lambda-m^{-1})%
\rvert=1$. Using that value of $n$, we define a scaled queue length process $%
Z(t)=(Z_1(t),Z_2(t))$ as follows: 
\begin{equation}
Z(t)=n^{-1/2}Q(nt)\text{ for }t\geq 0.  \label{eq:define:Z}
\end{equation}

In a stream of work including \citet{harrison1988brownian} and %
\citet{harrison1997dynamic}, the following two-stage scheme was developed
for approximating a queueing control problem in the heavy traffic parameter
regime. First, the original problem is approximated by
a \textit{Brownian control problem} (BCP) whose state descriptor $Z(t)$
corresponds to the scaled queue length process defined above. Second, the
BCP is replaced by a \textit{reduced Brownian control problem}, also called
the \textit{equivalent workload formulation} (EWF), whose state descriptor $%
W(t)$ corresponds to a certain linear transformation of $Z$ and is called a 
\textit{workload process}. For the simple example considered in this
section, $W$ is actually identical to $Z$, and there is no difference
between the original BCP and its EWF. To be specific, the BCP is a singular
control problem of the form specified in Section \ref{sec:general} (the formulation \textit{without} exogenous reflection at the boundary), with
drift vector 
\begin{equation*}
\xi =(\sqrt{n}(\lambda-m^{-1}),0)=(-1,0),
\end{equation*}%
covariance matrix 
\begin{equation*}
A=\left[ 
\begin{array}{cc}
2 & -1 \\ 
-1 & 2%
\end{array}%
\right] ,
\end{equation*}
control matrix 
\begin{equation*}
G=\left[ 
\begin{array}{cc}
1 & 0 \\ 
-1 & 1%
\end{array}%
\right],
\end{equation*}%
holding cost function 
\begin{equation*}
h(w)=h_1w_1+h_2w_2=w_1+2w_2,
\end{equation*}%
control cost vector $c=0$, and interest rate for discounting 
\begin{equation*}
\gamma=nr=4.
\end{equation*}

\begin{figure}[!ht]
\subfigure[Singular control policy]{
		\includegraphics[width=3.in]{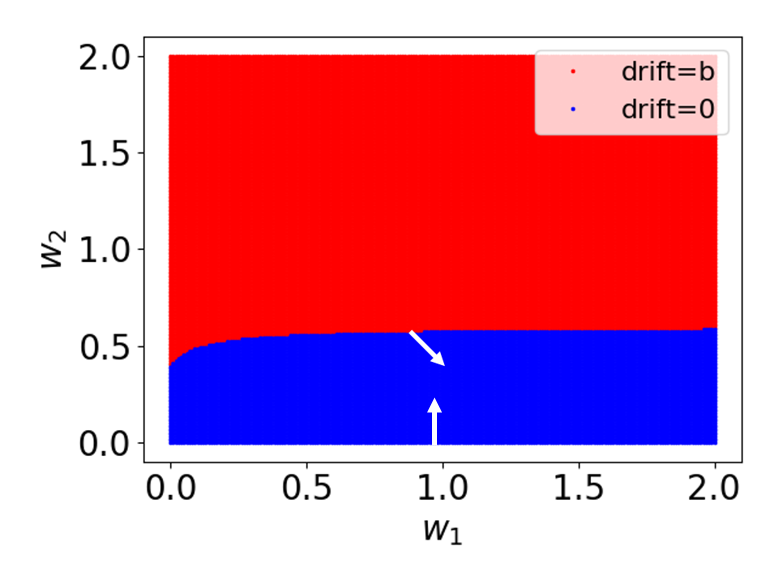}} \hfill 
\subfigure[Two policies for tandem queues]{
		\includegraphics[width=3.in]{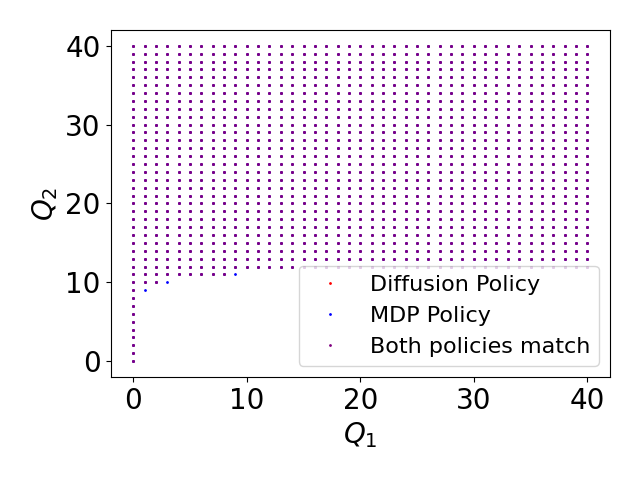}}
\caption{Optimal policies for the singular control and MDP formulations.}
\label{fig:2d:tandem}
\end{figure}

Applying our computational method to this singular control problem, we set
the upper bound on the feasible drift rates at $b=20$. The resulting optimal
policy has the bang-bang form shown in the left-hand panel of Figure \ref{fig:2d:tandem}: the control component $U_1$ (corresponding to idleness of server 1) increases at the maximum allowable rate in the red region of the state space, and it increases not at all in the complementary blue region, except when the vertical axis is hit; on the other hand, $U_2$ increases only when $W$ reaches the horizontal axis. In our singular control formulation, increases in $U_1$ produce an instantaneous displacement downward and to the right throughout the red region, whereas increases in $U_2$ produce instantaneous reflection in the vertical direction from the boundary $W_2=0$.

To interpret this solution in our original problem context, readers should
recall that state $(W_1,W_2)$ for the Brownian control problem corresponds to system state $(Q_1,Q_2)=(20W_1,20W_2)$ in the original tandem queueing network. 
The obvious interpretation of our computed solution is the following: server 2 is never idled except when it
has no work to do (that is, except when buffer 2 is empty), but server 1 is
idled if either it has no work to do or $W$ lies in the region colored red in
Figure \ref{fig:2d:tandem}; the exact computational meaning of this description will be spelled out in Appendix \ref{appendix:pre-limit:policies}.

In addition to solving the singular control problem, we have also solved the
Markov decision problem (MDP) associated with the original tandem queueing
network. The right-hand panel of Figure \ref{fig:2d:tandem} displays the
optimal policies derived from both the exact MDP formulation and the
approximating singular control formulation; the latter is referred to as the
``diffusion policy" in the figure. For the MDP solution, blue dots signify
states in which server 1 is idled, and for the diffusion policy, red dots
signify such states. Figure \ref{fig:2d:tandem} shows that the two polices
are strikingly close. Also, Table \ref{tab:simulation:tandem：2d} reports
the simulated performance (that is, the average present value of all holding
costs incurred) for the two policies applied in our original queueing
network, assuming both buffers are initially empty, along with standard errors. For comparison, Table \ref%
{tab:simulation:tandem：2d} also shows the performance of the policy that
never idles either server except when its buffer is empty. In our
simulations, one sees that the proposed policy derived from the singular
control approximation actually performs as well as the optimal policy from
the MDP formulation. 

\begin{table}[!htb]
\centering
\begin{tabular}{ccc}
\toprule Never idle & MDP & Diffusion \\ 
\midrule 1780 $\pm$ 1.0 & 1703 $\pm$ 0.9 & 1703 $\pm$ 0.9 \\ 
\bottomrule% &  & 
\end{tabular}%
\caption{Simulated performance for the tandem queues example.}
\label{tab:simulation:tandem：2d}
\end{table}

\section{The criss-cross network with various cost structures}\label{section:criss-cross}

\label{sec:criss-cross} 
\begin{figure}[ht]
\centering
\begin{tikzpicture}[start chain=going right,>=latex,node distance=1pt]
		% the rectangular shape with vertical lines
		\node[three sided,minimum width=1.5cm,minimum height = 1cm,on chain] (wa) {2};

		% the circle
		\node[draw,circle,on chain,minimum size=1cm] (se) {S1};
		\node[three sided2,minimum width=1cm,minimum height = 1.5cm,on chain]  (wa4)[above =of 
		se]  {1};
		
		\node[three sided,minimum width=1.5cm,minimum height = 1cm,on chain]  (wa5)[right =of 
		se, xshift=1cm]  {3};
		\node[draw,circle,on chain,minimum size=1cm] (se2) {S2};

		% the arrows and labels
		\draw[->] (se) edge node[above] {} (wa5.west);
		\draw[->] (se2.east) -- +(30pt,0);
		\draw[->] (se.south) -- +(0,-30pt);
		
		\draw[<-] (wa.west) -- +(-20pt,0) node[left] {$\lambda_2$};
		\draw[<-] (wa4.north) -- +(0,20pt) node[above] {$\lambda_1$};
		
	\end{tikzpicture}
\caption{The criss-cross network.}
\label{fig:criss-cross}
\end{figure}
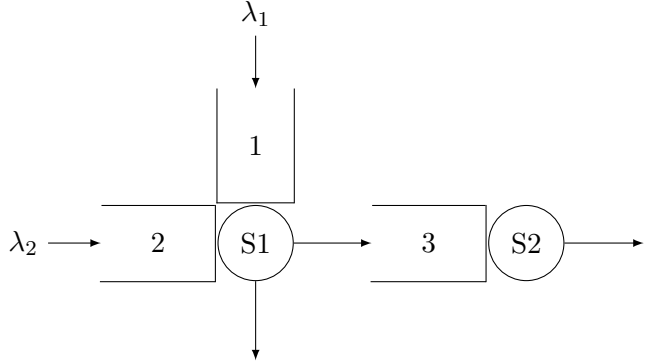

As shown in Figure \ref{fig:criss-cross}, the
criss-cross network consists of two single-server stations serving jobs of
three different classes. Jobs arrive to classes 1 and 2 according to
independent Poisson processes with rates $\lambda_1$ and $\lambda_2$,
respectively, and both of those classes are served at station 1. After
completing service, class 1 jobs leave the system, whereas class 2 jobs make
a transition to class 3 and are then served at station 2. As in Section \ref%
{sec:tandem}, we suppose that jobs of each class reside in a similarly
numbered buffer (represented by an open-ended rectangle in Figure \ref%
{fig:criss-cross}) both before and during their service.

Class $k$ jobs have
exponentially distributed service times with mean $m_k$ ($k=1,2,3$), and the
three service time sequences are mutually independent. Also, for maximum simplicity, we assume again that any service can be interrupted at any time and resumed later without any efficiency loss. A system manager
decides, at each point in time, which job class to serve at each station,
including the possibility of idleness for either server. Because we assume a
positive holding cost rate for each class, and because only class 3 jobs are
served at station 2, an optimal policy will obviously keep server 2 busy
serving class 3 whenever buffer 3 is non-empty. Thus the problem boils down
to deciding, at each point in time, whether to serve a class 1 job at
station 1, serve a class 2 job there, or idle server 1. 

The criss-cross network was introduced by \citet{harrison1989scheduling},
who restricted attention to one particular set of cost parameters and
focused on the heavy traffic regime. \citet{martins1996heavy} revisited the
same model, also focused on the heavy traffic regime, but allowed more
general cost parameters. The latter authors identified five mutually
exclusive and collectively exhaustive cases with respect to cost parameters, and they
offered conjectures about the form of an optimal policy in each case. More
will be said about this early work below.

As in Section \ref{sec:tandem}, we denote by $Q_{k}(t)$ the number of class $%
k$ jobs in the system at time $t\geq0$, or equivalently, the content of
buffer $k$ at time $t$ ($k=1,2,3$). Defining the system state $%
Q(t)=(Q_{1}(t),Q_{2}(t),Q_{3}(t))$, and denoting by $h=(h_{1},h_{2},h_{3})$
the vector of positive holding cost parameters, the instantaneous cost rate
incurred by the system manager at time $t$ is given by the inner product $h
\cdot Q(t)$. \cite{harrison1989scheduling} took the holding cost parameters
to be $h_{1}=h_{2}=h_{3}=1$, which means that the instantaneous cost rate in
every state equals the total number of jobs in the system. %
\citet{martins1996heavy} allowed any combination of positive holding cost
parameters, and they identified the following cases: 
\begin{align}
&\text{Case I: }(h_{2}-h_{3})/m_{2}\geq h_{1}/m_{1} \\
&\text{Case II: }(h_{2}-h_{3})/m_{2}<h_{1}/m_{1}  \label{criss-cross:caseII}
\\
&\text{Case IIA: } (\ref{criss-cross:caseII}) \text{ plus } h_2\geq h_3\text{
and } h_2/m_2\geq h_1/m_1 \\
&\text{Case IIB: } (\ref{criss-cross:caseII}) \text{ plus } h_2< h_3\text{
and } h_2/m_2\geq h_1/m_1 \\
&\text{Case IIC: } (\ref{criss-cross:caseII}) \text{ plus } h_2\geq h_3\text{
and } h_2/m_2< h_1/m_1 \\
&\text{Case IID: } (\ref{criss-cross:caseII}) \text{ plus } h_2< h_3\text{
and } h_2/m_2< h_1/m_1.
\end{align}
The simplest of these is Case I, which was studied by \citet{chen1994control}
and will not be discussed further here. Rather, we focus attention on Case
II and numerically solve examples covering its four sub-cases.

Prior to that analysis, a few more words about existing literature are
appropriate. The specific cost structure studied by %
\citet{harrison1989scheduling} is an example of Case IIA. Those authors
proposed an effective policy for that case, and \citet{martins1996heavy}
proved the asymptotic optimality of that policy; also see \citet{budhiraja2005large}. \citet{martins1996heavy}
further made conjectures about the remaining sub-cases, two of which (Cases IIB and IIC) were
rigorously justified in later papers by %
\citet{budhiraja2008optimal} and \citet{budhiraja2017construction}.

Proceeding now to the analysis of numerical examples, we shall analyze the
four holding cost combinations specified in Table \ref%
{tab:criss-cross:parameter}. The stochastic parameters in all four numerical
examples will be $\lambda _{1}=1,\lambda _{2}=0.95$, $m_{1}=m_{2}=0.5$ and $%
m_{3}=1$. % The load factor for each server is then 0.95, meaning that 95 percent of each server's time is required to process all arrivals. 
With
these data we are in the heavy traffic regime, and as in Section \ref%
{sec:tandem}, a reasonable choice of the sequence index or scaling parameter
is $n=400$. Finally, the interest rate for discounting is again taken to be $%
r=0.01.$ 
\begin{table}[!htb]
\centering
\begin{tabular}{lccc}
\toprule & $h_1$ & $h_2$ & $h_3$ \\ 
\midrule IIA & 1 & 1 & 1 \\ 
IIB & 1 & 1 & 1.5 \\ 
IIC & 1.5 & 1 & 1 \\ 
IID & 1.5 & 1 & 1.5 \\ 
\bottomrule% &  &  & 
\end{tabular}%
\caption{Holding cost parameters for different cases.}
\label{tab:criss-cross:parameter}
\end{table}

Reference was made in Section \ref{sec:tandem} to a two-stage procedure for
deriving a heavy traffic approximation to a queueing network control
problem. For the criss-cross network, that analysis was undertaken by %
\citet{harrison1989scheduling}, as follows. First, a Brownian control
problem (BCP) is formulated whose state vector $Z(t)$ corresponds to the
three-dimensional scaled queue length process defined via (\ref{eq:define:Z}%
) as in Section \ref{sec:tandem}. Then an equivalent workload formulation
(EWF) is derived whose state vector $W(t)$ is defined for the criss-cross
network as 
\begin{equation}
W(t)=MZ(t)\text{ for }t\geq 0,  \label{eq:define:W}
\end{equation}%
where 
\begin{equation*}
M=\left[ 
\begin{array}{ccc}
0.5 & 0.5 & 0 \\ 
0 & 1 & 1%
\end{array}%
\right] .
\end{equation*}%
Obviously, the process $W$ inherits from $Z$ the scaling of time by a factor
of $n$ and the scaling of queue lengths by a factor of $\sqrt{n}$. We call $%
M $ a \textit{workload profile matrix}, interpreting $M_{kj}$ as the
expected amount of time that server $k$ must spend to complete the
processing of a job currently in class $j$. Thus, except for the scale
factors referred to above, one interprets $W_{k}(t)$ as the expected amount
of time required from server $k$ to complete the processing of all jobs
present anywhere in the system at time $t$ ($k=1,2)$. The control process $%
U(t)$ for the EWF is also two-dimensional, with $U_{k}(t)$ interpreted as a
scaled version of the cumulative idleness experienced by server $k$ up to
time $t$ ($k=1,2)$.

The EWF derived by \citet{harrison1989scheduling} for the criss-cross
network is a singular control problem of the form (\ref{problem:state:w})-(\ref{min:objective:new}) specified in Section \ref%
{sec:general} (again we use the formulation \textit{without} exogenous reflection at the boundary), with state dimension $d=2$, control dimension $p=2$, drift
vector 
\begin{equation*}
\xi =\sqrt{n} \text{ }(\lambda_1 m_1+\lambda_2 m_2-1,\text{ } \lambda_2 m_3
-1)=(-0.5,-1),
\end{equation*}%
covariance matrix 
\begin{equation*}
A=\left[ 
\begin{array}{cc}
1 & 0.5 \\ 
0.5 & 2%
\end{array}%
\right] ,
\end{equation*}
control matrix 
\begin{equation*}
G=\left[ 
\begin{array}{cc}
1 & 0  \\ 
0 & 1  \\ 
\end{array}%
\right],
\end{equation*}%
control cost vector $c=0$, and interest rate for discounting 
\begin{equation*}
\gamma=nr=4.
\end{equation*}%
Finally, the holding cost function $h(w)$ for the EWF is 
\begin{equation}
h(w)=\text{min}\{ h\cdot z: Mz=w, z \geq 0 \},  \label{eq:criss-cross cost}
\end{equation}
which specializes as in Table \ref{tab:criss-cross:cost function} for the
stochastic parameters assumed earlier and the holding cost parameters
specified in Table \ref{tab:criss-cross:parameter}. 
\begin{table}[!htb]
\centering
\begin{tabular}{lccc}
\toprule & If $w_{2}\geq 2w_{1}$ &  & If $w_{2}<2w_{1}$ \\ 
\midrule IIA & $w_2$ &  & 2$w_1$ \\ 
IIB & $-w_1+1.5w_2$ &  & 2$w_1$ \\ 
IIC & $w_2$ &  & $3w_1-0.5w_2$ \\ 
IID & $-w_1+1.5w_2$ &  & $3w_1-0.5w_2$ \\ 
\bottomrule% &  &  & 
\end{tabular}%
\caption{Cost function $h(w_{1},w_{2})$ for numerical values in Table 
\protect\ref{tab:criss-cross:parameter}.}
\label{tab:criss-cross:cost function}
\end{table}

An intuitive explanation of (\ref{eq:criss-cross cost}) is as follows. In
the heavy traffic regime, a system manager can navigate quickly (that is,
instantaneously in the heavy traffic limit) and costlessly between any two
(scaled) queue length vectors $z$ that yield the same (scaled) workload
vector $w$. In our criss-cross model, such navigation is accomplished by
adjusting service priorities at station 1: giving priority there to class 1
drives $z_1$ quickly down and $z_2$ quickly up in equal amounts, and
vice-versa, without affecting $w$. Thus, the system manager can dynamically
adjust priorities at station 1 so as to hold, at each point in time $t$, the
least costly queue length vector $Z(t)$ that is consistent with the workload
vector $W(t)$ that then prevails.

To review, heavy traffic analysis of the criss-cross network begins by
formulating a three dimensional Brownian control problem (BCP) whose state
vector $Z(t)$ is a scaled version of the model's three-dimensional queue
length process $Q(t)$. Then the BCP is reduced to a two-dimensional
equivalent workload formulation (EWF) whose state vector $W(t)$ is defined
in terms of $Z(t)$ via (\ref{eq:define:W}), and the non-linear cost function
for the EWF is given by (\ref{eq:criss-cross cost}). Actually, there are
four different versions of the criss-cross EWF, using the four different
cost structures specified in Table \ref{tab:criss-cross:cost function}. We have
applied our computational method to all four of those singular control
problems, setting the algorithm's upper bound on drift rates at $b=20$ as in
our previous analysis of tandem queues in Section \ref{sec:tandem}.

The resulting solutions are displayed graphically in Figure \ref%
{fig:4cases:criss-cross}, where displacement to the right corresponds to
idleness of server 1, and upward displacement corresponds to idleness of server 2. In
Case IIA, each server is idled only when there is no work for it to do
anywhere in the system. In the other three cases, our solution applies
maximal drift to the right, interpreted as idleness of server 1, in
the upper red region if there is one, and applies maximal upward drift,
interpreted as idleness of server 2, in the lower red region if
there is one. On the other hand, the computed solution applies no drift in
the blue region except at the axes, which is interpreted to mean that each
server continues to work at full capacity in those regions except when there
is no work for that server anywhere in the system.

\begin{figure}[!ht]
\subfigure[IIA]{
		\includegraphics[width=3.in]{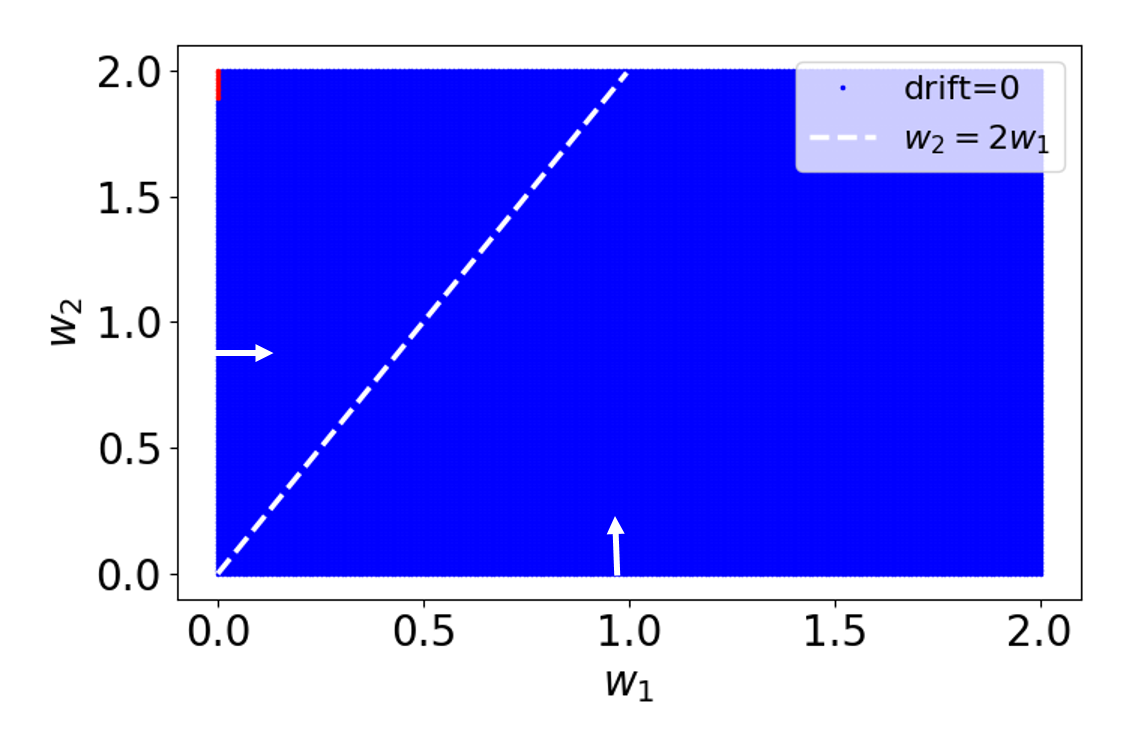}} \hfill 
\subfigure[IIB]{
		\includegraphics[width=3.in]{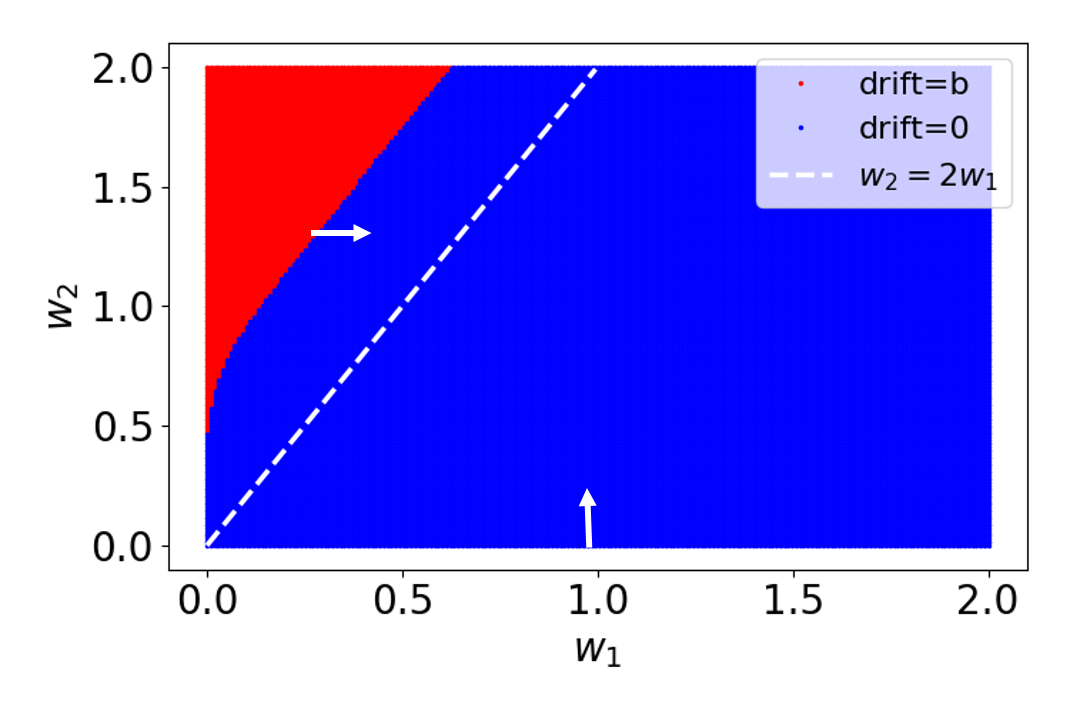}} 
\subfigure[IIC]{
		\includegraphics[width=3.in]{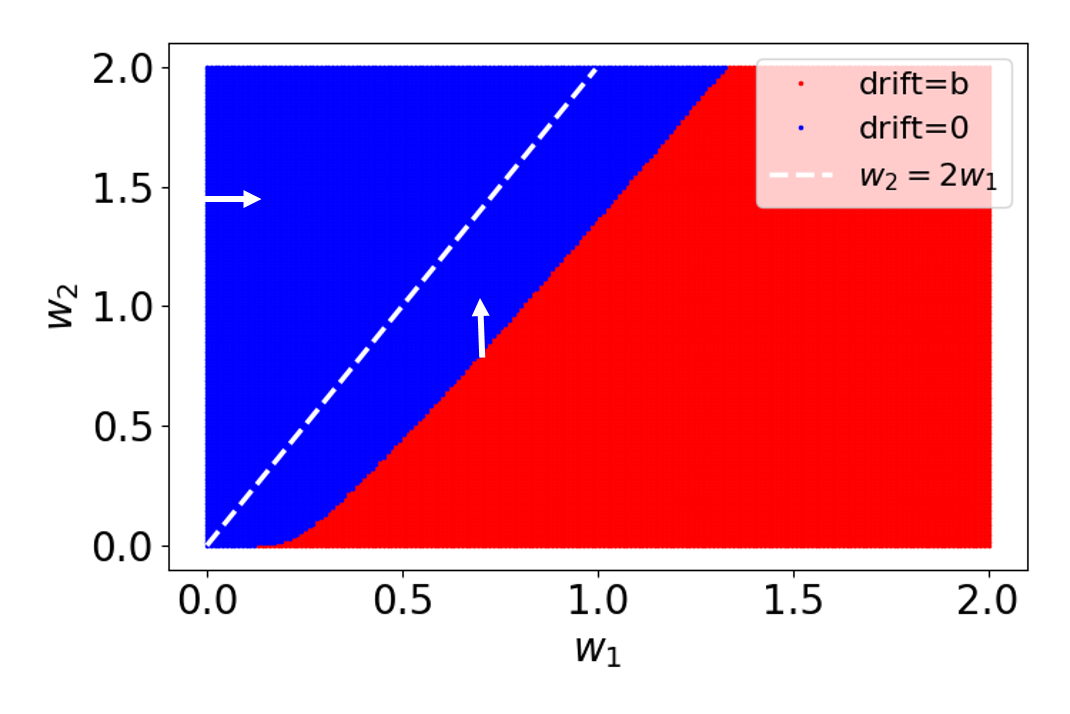}} \hfill 
\subfigure[IID]{
		\includegraphics[width=3.in]{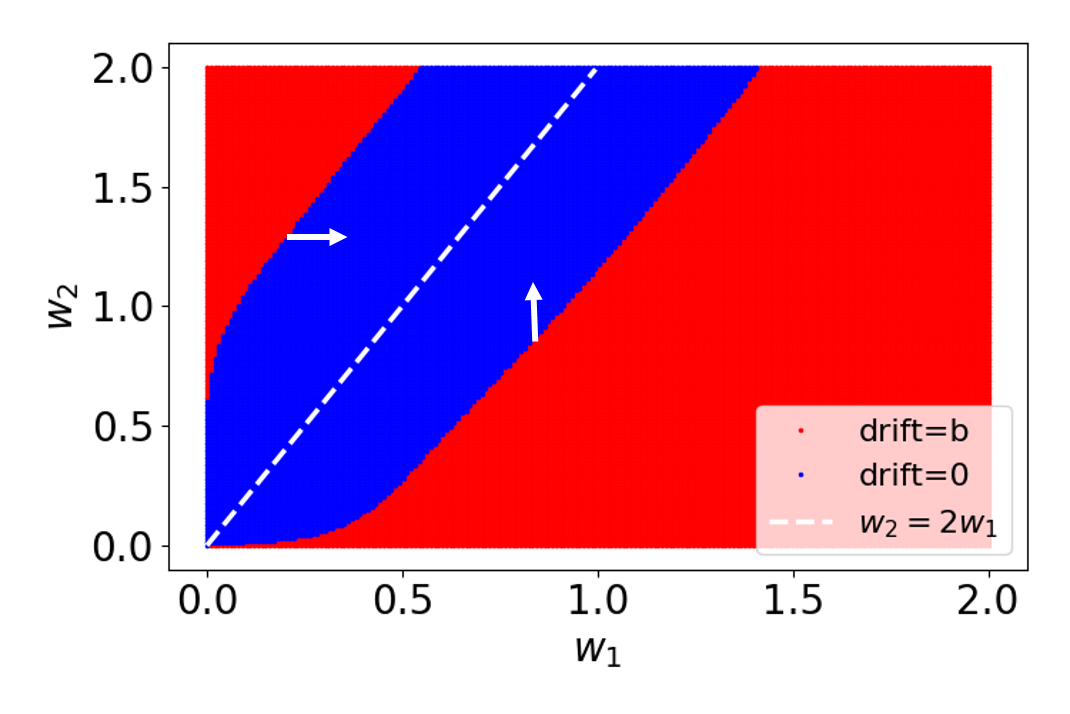}}
\caption{Graphical representation of the diffusion polices derived via solution of the approximating singular control problem.}
\label{fig:4cases:criss-cross}
\end{figure}

To understand why and how these idleness prescriptions are to be
accomplished, one must consider the rest of the computed solution, that is,
the scaled queue length vector $z$ in which the scaled workload $w$ is to be
held. Equations (8.5a) and (8.5b) of \citet{martins1996heavy} show that the
minimum in (\ref{eq:criss-cross cost}) is achieved in all four of our cases by 
\begin{equation}
z^{*}(w)=((2w_1 - w_2 )^{+}, \, 2w_1\wedge w_2,\, (w_2-2w_1)^{+}).  \label{eq:opt z}
\end{equation}
Superficially, this appears to say, in Case IIA for example, that buffer 3 should remain
empty whenever $W$ falls below the dotted white line in Figure 
\ref{fig:4cases:criss-cross}, even though server 2 is to be idled only when $W$
reaches the horizontal axis. To resolve this seeming contradiction, %
\citet{harrison1989scheduling} interpreted (\ref{eq:opt z}) to mean the
following: class 1 (the class that exits the system when its one service is
completed) should be given priority at station 1 whenever $Q_1(t)>0$ and $%
Q_3(t)>s$ (mnemonic for \textit{safefy stock}), where $s>0$ is a tuning parameter to be optimized via simulation;
otherwise Class 2 should be given priority. The intuitive argument
supporting this policy is that granting default priority to class 1 greedily
reduces the instantaneous cost rate incurred by the system manager, but by
taking $s$ large enough one can still protect server 2 from starvation.

\citet{martins1996heavy} generalized the policy proposed by %
\citet{harrison1989scheduling} to cover all four cases as follows. First,
idle server 1 whenever $W$ reaches the vertical axis or falls in the
upper red region in Figure \ref{fig:4cases:criss-cross}. Second, idle server
2 whenever $W$ reaches the horizontal axis or falls in the lower red
region in Figure \ref{fig:4cases:criss-cross}; this happens through prioritizing
class 1 at server 1, which starves server 2. Third, give class 1 priority
at station 1 if $W$ falls in the blue region and $Q_3>s$;
otherwise, give priority to class 2. Here again, $s>0$ is a tuning parameter to
be optimized via simulation, and readers may consult Appendix \ref{appendix:pre-limit:policies} for the exact computational meaning of the verbal policy description given here. Hereafter, our policy derived from the singular control approximation will be referred to as the 
\textit{diffusion policy} for the criss-cross network.

\begin{figure}[!ht]
\subfigure[IIB, $Q_1=0$]{
			\includegraphics[width=3.in]{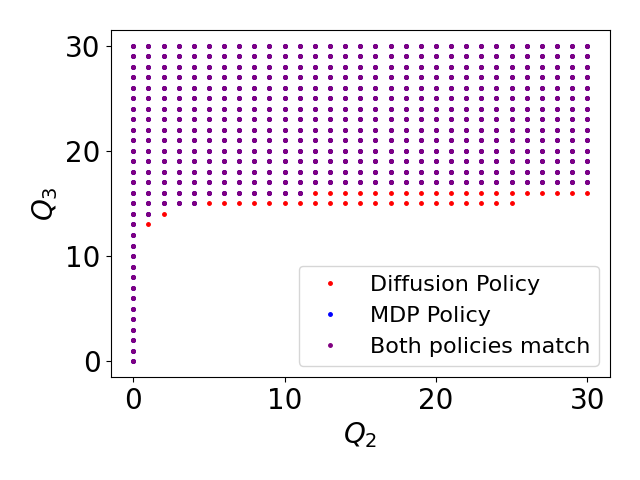}}
\hfill 
\subfigure[IIC, $Q_3=0$]{
			\includegraphics[width=3.in]{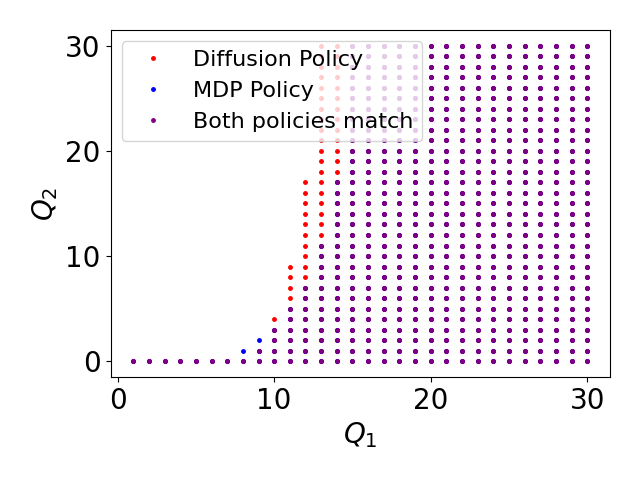}}
\subfigure[IID-1, $Q_1=0$]{
			\includegraphics[width=3.in]{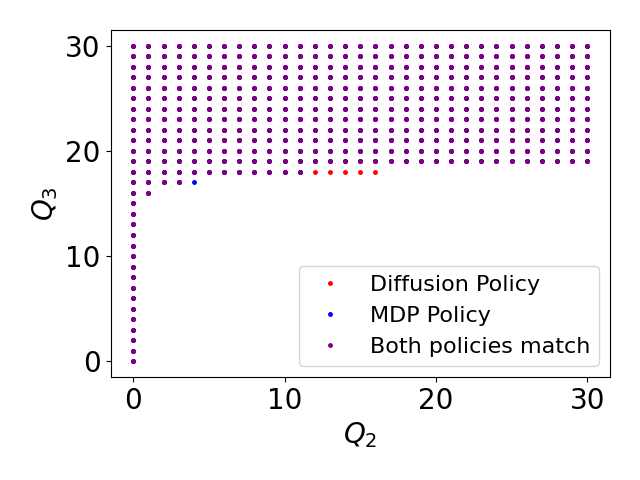}}
\hfill 
\subfigure[IID-2, $Q_3=0$]{
			\includegraphics[width=3.in]{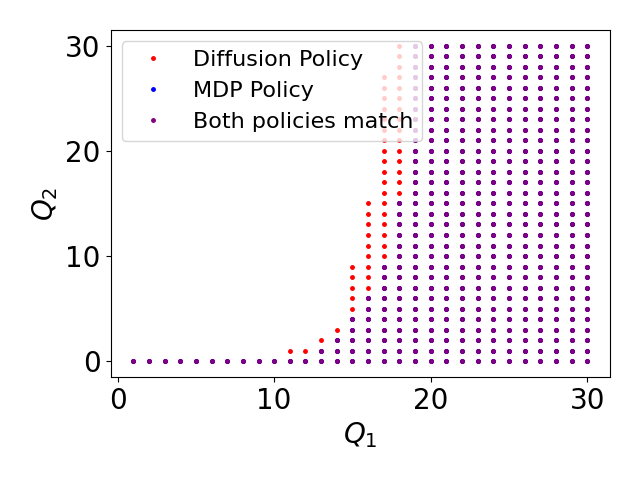}}
\caption{Graphical representation of two polices for the criss-cross example: the optimal MDP policy versus our proposed policy based on the singular control approximation.}
\label{fig:mdp:criss-cross}
\end{figure}

For comparison, we solved the associated MDP numerically to obtain an exact
optimal policy. Figure \ref{fig:mdp:criss-cross} displays the optimal MDP policy. Motivated by (\ref{eq:opt z}), 
one expects $Q_1(t) \approx 0$ whenever the workload vector is above the dotted white line in Figure \ref{fig:4cases:criss-cross}. Thus, we restrict attention to the case $Q_1 = 0$ in panels (a) and (c) of Figure \ref{fig:mdp:criss-cross} and display the optimal action as a function of $Q_2$ and $Q_3$. For the MDP policy, blue dots signify states in which server 1 is idled, and for our diffusion policy derived from the approximating singular control problem (see Algorithm \ref{algo:pre-limit:criss-cross} in Appendix \ref{appendix:pre-limit:policies}), red dots signify such states. Similarly, one expects $Q_3(t) \approx 0$ whenever the workload vector falls below the dotted white line. Therefore, we restrict attention to the case $Q_3 = 0$ in panels (b) and (d) of Figure \ref{fig:mdp:criss-cross} and display the optimal action as a function of $Q_1$ and $Q_2$. For the MDP policy, blue dots signify states in which server 1 prioritizes class 1, which starves server 2 and causes it to idle, and for the diffusion policy, red dots signify such states. Figure \ref{fig:mdp:criss-cross} shows that the two policies are very similar. 
\begin{table}[!htb]
\centering

\begin{tabular}{lcccc}
\toprule Cases & Static priority &  & MDP & Our proposed policy \\ 
\midrule IIA & 1765 $\pm$ 1.1 &  & 1488 $\pm$ 0.9 & 1491 $\pm$ 0.9 $(s^* =2)$
\\ 
IIB & 2084 $\pm$ 1.1 &  & 1659 $\pm$ 1.0 & 1665 $\pm$ 1.0 $(s^* = 2)$ \\ 
IIC & 1812 $\pm$ 1.0 &  & 1684 $\pm$ 1.1 & 1690 $\pm$ 1.0 $(s^* = 1)$ \\ 
IID & 2134 $\pm$ 1.1 &  & 1872 $\pm$ 1.0 & 1880 $\pm$ 1.1 $(s^* = 1)$ \\ 
\bottomrule  % &  &  &  & 
\end{tabular}%
\caption{Simulated performance of three policies for cases IIA - IID.}
\label{tab:simulation:criss-cross}
\end{table}

In addition, Table \ref{tab:simulation:criss-cross} reports the simulated
performance (that is, the average present value of all holding costs
incurred), with standard errors, of three policies for the original
criss-cross example: i) the policy that gives static priority to class 1 everywhere in
the state space, ii) the diffusion policy derived via solution of the
approximating singular control problem, and iii) the MDP optimal policy. The
performance of the diffusion policy is close to that of the optimal MDP
policy (within a fraction of one percent) and much better than that of the
static priority policy.

\section{Several variants of a three-station example}

\label{sec:BIGSTEP example}

\begin{figure}[ht]
\centering
\includegraphics[width=6in]{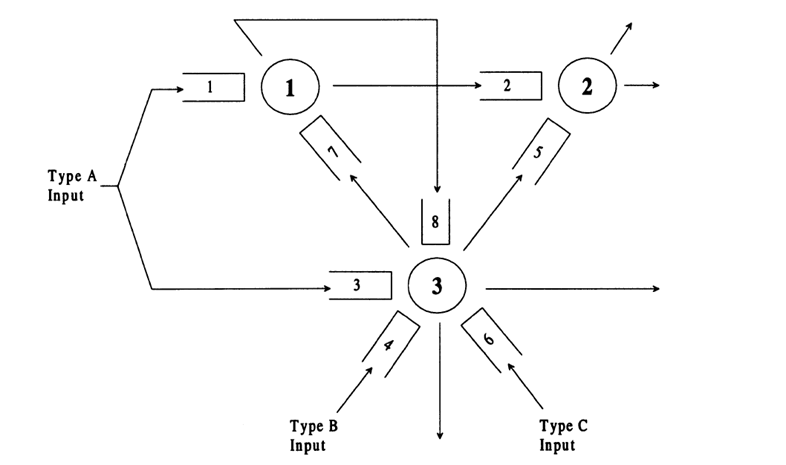}
\caption{Picture of the 3-station queueing network (copied from 
\citet{harrison1996bigstep}).}
\label{fig:three-station}
\end{figure}

\citet{harrison1996bigstep} introduced the three-station queueing network
displayed in Figure \ref{fig:three-station}, which was further studied by %
\citet{harrison1997dynamic}. It has three external input processes for jobs
of types A, B and C. Jobs of types B and C follow the predetermined routes
shown in Figure \ref{fig:three-station}. For jobs of type A, there are two
processing modes available, and the routing decision must be made
irrevocably upon arrival. If a type A job is directed to the lower route, it
requires just a single service at station 3, but if it is directed to the
upper route, then two services are required at stations 1 and 2, as shown in
Figure \ref{fig:three-station}.

We define a different job class for each combination of job type and stage
of completion. Jobs of each class wait for service in a designated buffer,
represented by the open-ended rectangles in Figure \ref{fig:three-station}.
There are eight classes in total, and each station serves multiple classes.
Thus, the system manager must make dynamic sequencing decisions. She also
makes dynamic routing decisions to determine which route to use for each
type A job. In addition to the dynamic routing and sequencing decisions, the
system manager can reject new jobs of any type upon arrival. On the one
hand, she incurs a penalty for each rejected job, and on the other hand,
there are linear holding costs that provide a motivation for rejecting new
arrivals when queues are large. We denote by $h_1, \ldots, h_8$ the holding
cost rates (expressed in units like dollars per hour) for the eight job
classes identified in Figure \ref{fig:three-station}.

Jobs of types A, B, and C arrive to the system according to independent
Poisson processes with rates 0.5, 0.25 and 0.25 jobs per hour, respectively.
In addition, class $k$ jobs have exponentially distributed service times
with mean $m_k$ where 
\begin{equation*}
( m_1, \ldots, m_8 ) = (2,3,1,1,1,1,2,1).
\end{equation*}
As an initial routing mechanism, to be modified by the system manager's
dynamic control actions, we suppose that routing decisions for type A
arrivals are made by independent coin flips, resulting in the following
vector of average external arrival rates into the various classes: 
\begin{equation*}
( \lambda_1, \ldots, \lambda_8 ) =\left ( \frac{1}{4}, 0, \frac{1}{4}, \frac{%
1}{4}, 0, \frac{1}{4}, 0 ,0 \right).
\end{equation*}%

\begin{figure}[tbp]
\centering
\includegraphics[width=6in]{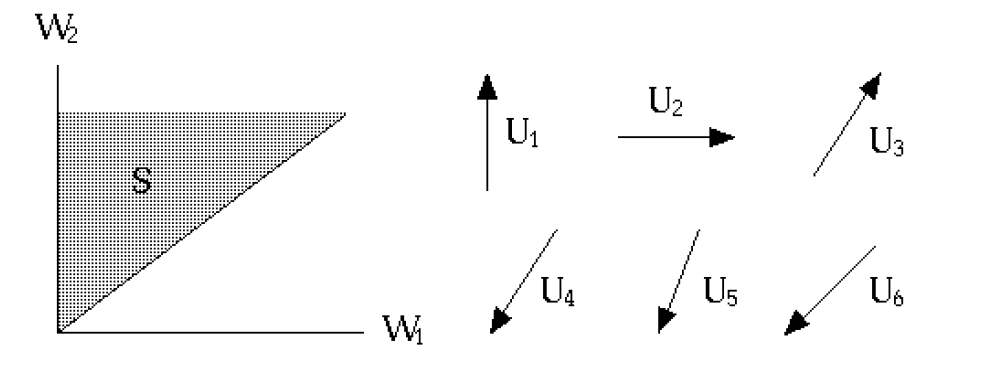}
\caption{The set $S$ and directions of available controls (copied from 
\citet{harrison1996bigstep}).}
\label{fig:bigstep:control}
\end{figure}

\citet{harrison1996bigstep} and \citet{harrison1997dynamic} derived the
approximating Brownian control problem and its equivalent workload
formulation (EWF) for the three-station queueing network. The EWF will be
recapped in this section and then solved numerically for several different
cost structures. It is a singular control problem \textit{without} exogenous reflection at the boundary, defined via (\ref{problem:state:w})- (\ref{min:objective:new}) in Section \ref{sec:general},
except that its state space is not an orthant. Rather, it is the wedge 
\begin{equation*}
S = \{w \in \mathbb{R}^2 : w = Mz, z \geq 0\}
\end{equation*}
pictured in Figure \ref{fig:bigstep:control}, where $M$ is the $2 \times 8$
workload profile matrix 
\begin{equation*}
M=\left[ 
\begin{array}{cccccccc}
2 & 0 & 2 & 2 & 0 & 6 & 4 & 2 \\ 
3 & 3 & 3 & 4 & 1 & 6 & 3 & 3%
\end{array}%
\right] .
\end{equation*}
As in Section \ref{sec:criss-cross}, the workload process $W$ for our
three-station example is defined by the relationship $W(t)=MZ(t),\mathit{\ }
t \geq 0$, where $Z$ is a scaled queue length process with one component for
each job class or buffer, so $Z$ is eight dimensional for the current
example. In Section \ref{sec:criss-cross} we saw that the workload dimension
of the criss-cross network (that is, the dimension of its EWF) equals the
number of that system's service stations, but in the current example the
system manager's dynamic routing capability reduces the workload dimension
from three (the number of service stations) to two. %
\citet{harrison1997dynamic} interpret components of $W$ as scaled workloads
for two overlapping ``server pools," one composed of servers 1 and 3, the
other composed of servers 2 and 3.

In addition to the state space $S$, Figure \ref{fig:bigstep:control} shows
the six directions of control that are available by increasing different
components of the control process $U$. Those six directions of control are
the columns of the matrix 
\begin{equation*}
G=\left[ 
\begin{array}{cccccc}
1 & 0 & 2 & -1 & -1/2 & -3/2 \\ 
0 & 1 & 3 & -3/2 & -1 & -3/2%
\end{array}%
\right] .
\end{equation*}
Control modes 1 through 3 correspond to idling servers 1 through 3, while 
control modes 4, 5 and 6 correspond to rejecting arrivals of types A, B and
C, respectively. The associated control cost vector $c$ has the form $c =
(0, 0, 0, c_4, c_5, c_6)$, where $c_4, c_5 \text{ and } c_6$ will be given
strictly positive values in the numerical examples to follow. The three zero
components of $c$ reflect an assumption that there is no direct cost of
idling servers, whereas the positive values of $c_4, c_5 \text{ and } c_6$
represent the costs (expressed in units like dollars per job) to reject new
arrivals of types A, B and C, respectively; one can think of those ``costs"
either as direct penalties or as opportunity costs for business foregone. It
is noteworthy that, because the third column of $G$ can be written as a
positive linear combination of the first and second columns, and
furthermore, control modes 1, 2 and 3 are all costless (that is, $%
c_1=c_2=c_3=0$), control mode 3 can actually be eliminated from the problem
formulation. Equivalently, one can set $U_3(t)=0$ for all $t \geq 0$ without
loss of optimality, which is consistent with the optimal policies derived
below.

The final datum for the exact version of our three-station
network control problem is the following vector of holding cost parameters
for the eight job classes: 
\begin{equation*}
h=(h_1, \ldots, h_8) = (6,3,6,6,1,12,7,6).
\end{equation*}
We now define the holding cost function $h(\cdot)$ for our EWF as in Section %
\ref{sec:criss-cross}, that is, 
\begin{equation}
h(w)=\min\{h \cdot z:Mz=w,z\geq 0 \},\quad w\in S,
\end{equation}
\label{eqn:def:running-cost:3station-example} and with the holding cost
parameters specified above, it can be verified that (\ref%
{eqn:def:running-cost:3station-example}) is equivalent to 
\begin{equation}
h(w) = w_1 + w_2, \quad w \in S.
\end{equation}
\label{eq:cost function}

Let us now consider the singular control problem that is the EWF for our
three-station example. Its state space $S$, control matrix $G$, holding cost
function $h(w)$, and control cost vector $c$ are as specified above. Next,
the underlying Brownian motion $X=\{X(t), t\geq 0\}$ that appears in
equations (\ref{problem:state:w})-(\ref{problem:U}) is two-dimensional, with
drift vector $\xi=[-5.0,-5.0]$ and covariance matrix 
\begin{equation*}
A = \left[ 
\begin{array}{cc}
50 & 54 \\ 
54 & 69%
\end{array}%
\right] .
\end{equation*}%
Finally, we take the interest rate for discounting in the EWF to be $%
\gamma=0.1$.

To formulate or specify the drift control problem by which we approximate
the EWF (see Section \ref{sec:drift}), observe that the first two columns of
the control matrix $G$ form the $2 \times 2$ submatrix 
\begin{equation*}
R = 
\begin{bmatrix}
1 & 0 \\ 
0 & 1%
\end{bmatrix}%
,
\end{equation*}
which satisfies assumption (\ref{submatrix assumption}) of our singular control formulation. To accommodate the non-orthant state space of our current example, we must take care to specify which column of $R$ provides the ``back-up reflection capability" (see Section \ref{sec:drift} for an explanation of that phrase) at each of the rays that form the boundary of $S$. For that purpose, we associate the first
column of $R$ with the vertical axis in Figure \ref{fig:bigstep:control} and associate its second column with the oblique ray that forms the other boundary of $S$. It is crucial, of course, that each
column of $R$ points into the interior of $S$ from the boundary ray with which it is associated. In fact, it can be verified that each of the two columns of $R$ is the \textit{unique} column in $G$ that points into the interior of $S$ from its corresponding boundary ray. (Recall that we set $U_3 \equiv 0$ without loss of optimality.)

\begin{figure}[!ht]
\subfigure[\citet{kushner1991numerical}]{
		\includegraphics[width=3.in]{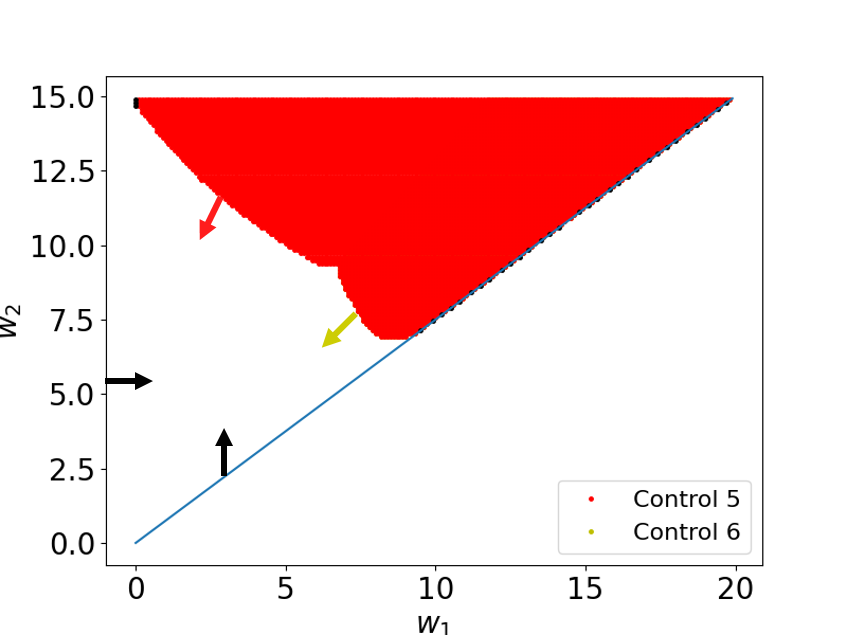}} \hfill 
\subfigure[Proposed policy]{
		\includegraphics[width=3.in]{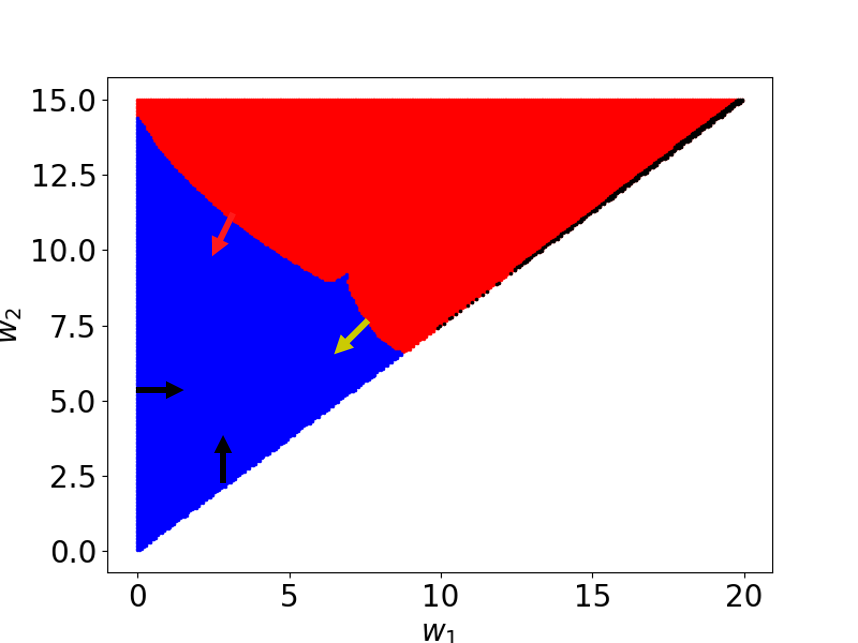}}
\caption{Graphical representation of the policies in case 2.}
\label{fig:bigstep:2a}
\end{figure}

\begin{figure}[!ht]
\subfigure[\citet{kushner1991numerical}]{
		\includegraphics[width=3.in]{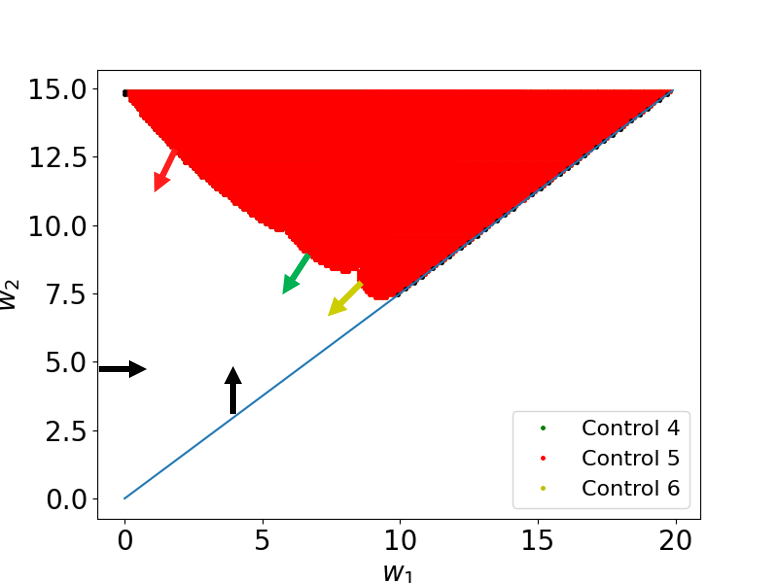}} \hfill 
\subfigure[Proposed policy]{
		\includegraphics[width=3.in]{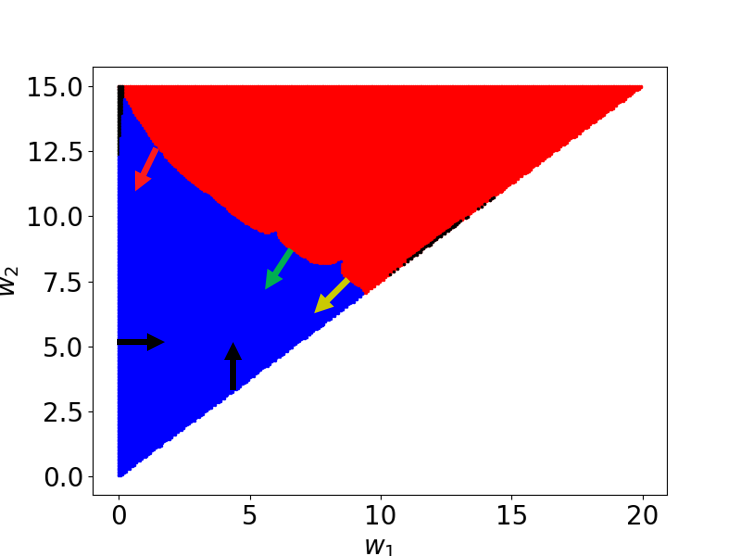}}
\caption{Graphical representation of the policies in case 3.}
\label{fig:bigstep:3}
\end{figure}

After making minor changes in the code that implements our computational
method, so that it treats a problem in the wedge rather than the quadrant,
and setting the upper bound on allowable drift rates to $b=200$, we
solved the singular control problem for the three combinations of cost
parameters specified in Table \ref{tab:bigstep:cost}. 
\begin{table}[ht]
\centering
\begin{tabular}{lccc}
\toprule & $c_4$ & $c_5$ & $c_6$ \\ 
\midrule Case 1 & 2 & 1 & 1 \\ 
Case 2 & 2 & 1 & 2 \\ 
Case 3 & 1.65 & 1 & 2.25 \\ 
\bottomrule% &  &  & 
\end{tabular}%
\caption{The cost parameters.}
\label{tab:bigstep:cost}
\end{table}
In addition, the same three singular control problems were solved using the
method proposed by \citet{kushner1991numerical}. See Figures \ref%
{fig:bigstep:1c}, \ref{fig:bigstep:2a}, and \ref{fig:bigstep:3} for
graphical comparisons of the policies derived with the two methods in Cases
1, 2 and 3, respectively.

The solutions produced by the two computational methods are very similar
visually, and they share the following gross properties: (i) in all three
cases, servers 1 and 2 are idled only when there is no work for them
anywhere in the system, and server 3 is never idled; (ii) in Case 1, new
arrivals of type C are rejected under certain circumstances, but those of
types A and B never are; (iii) in Case 2, new arrivals of types B and C are
rejected in different circumstances, but those of type A never are; and in
Case 3, new arrivals of all three types are rejected in different
circumstances. 
\begin{figure}[H]
\subfigure[\citet{kushner1991numerical}]{
		\includegraphics[width=3.in]{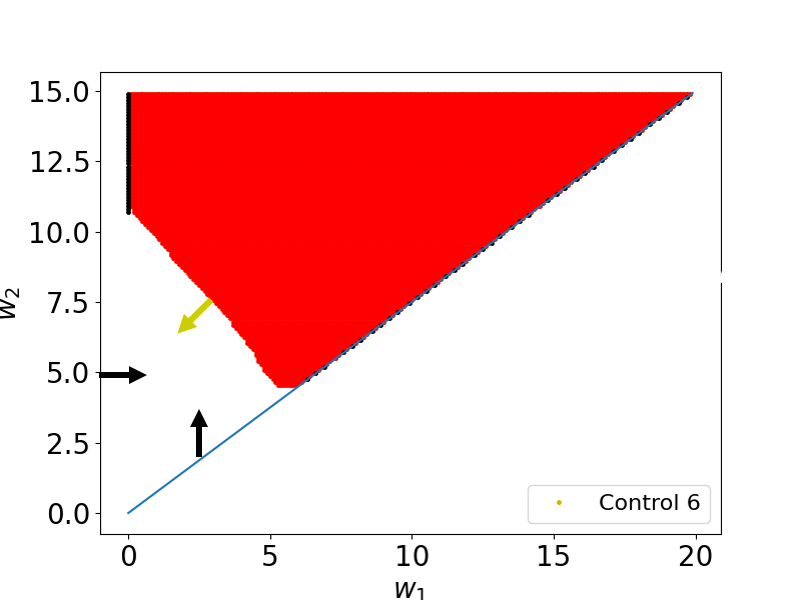}} \hfill 
\subfigure[Proposed policy]{
		\includegraphics[width=3.in]{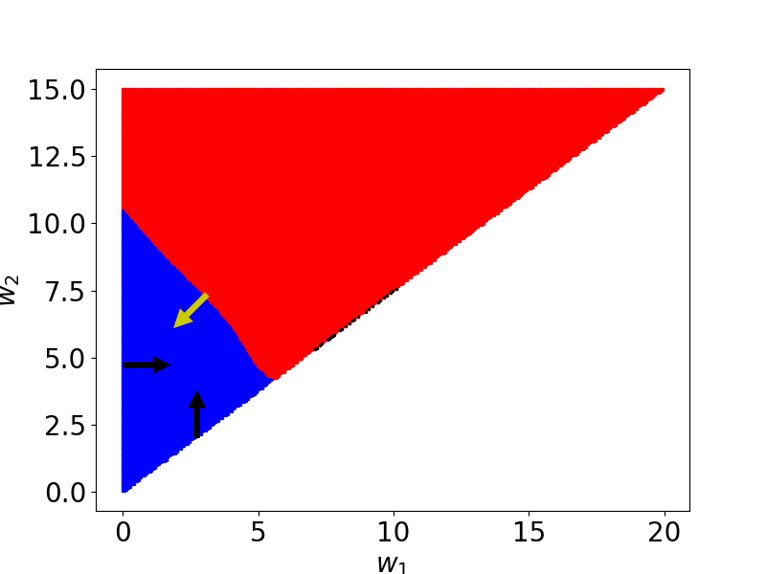}}
\caption{Graphical representation of the policies in case 1.}
\label{fig:bigstep:1c}
\end{figure}

Finally, Table \ref{tab:simulation:bigstep：2d} reports the simulated
performance (that is, the average present value of all costs incurred) in
all three cases for policies derived via the two methods, assuming both buffers are initially empty, along with the
associated standard errors. For comparison, Table \ref%
{tab:simulation:bigstep：2d} also shows the performance of the policy that
never turns away jobs and never idles any server except when such idleness
is unavoidable. It should be emphasized that these performance estimates
were derived by simulating sample paths \textit{of the Brownian singular
control problem itself}, not by simulating performance in our original
queueing network, as was the case in Sections \ref{sec:tandem} and \ref%
{sec:criss-cross}. The reason for this is that, for our three-station
network example, it is not obvious how to interpret or implement the EWF
solution in our original problem context, so we postpone that task to future
research. See Section \ref{sec:conclude} for further discussion. 
\begin{table}[htb]
\centering
\begin{tabular}{lccc}
\toprule & No control & Proposed policy & \citet{kushner1991numerical} \\ 
\midrule Case 1 & 102.7 $\pm$ 0.12 & 67.7 $\pm$ 0.07 & 67.7 $\pm$ 0.07 \\ 
Case 2 & 102.7 $\pm$ 0.12 & 84.2 $\pm$ 0.10 & 84.2 $\pm$ 0.10 \\ 
Case 3 & 102.7 $\pm$ 0.12 & 85.2 $\pm$ 0.07 & 85.1 $\pm$ 0.07 \\ 
\bottomrule% &  &  & 
\end{tabular}%
\caption{Simulation performance for three class of a three-station example.}
\label{tab:simulation:bigstep：2d}
\end{table}

\section{Many queues in series}
\label{Sec:Many queues}
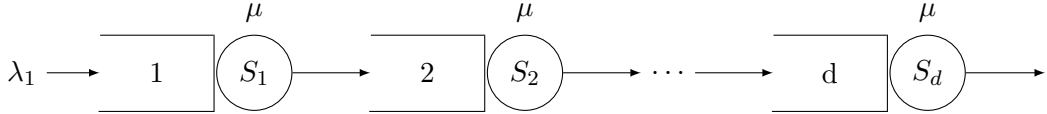
\begin{figure}[!htb]
\centering
\begin{tikzpicture}[start chain=going right,>=latex,node distance=1pt]
		% the rectangular shape with vertical lines
		\node[three sided,minimum width=1.5cm,minimum height = 1cm,on chain] (wa) {1};

		% the circle
		\node[draw,circle,on chain,minimum size=1cm] (se) {$S_1$};

		\node[three sided,minimum width=1.5cm,minimum height = 1cm,on chain]  (wa5)[right =of 
		se, xshift=1cm]  {2};
		\node[draw,circle,on chain,minimum size=1cm] (se2) {$S_2$};
		\node[] (wadot)[right =of  se2, xshift=1cm]  {$\ldots$};
		
		\node[three sided,minimum width=1.5cm,minimum height = 1cm,on chain]  (wad)[right =of  wadot, xshift=1cm]  {d};
		\node[draw,circle,on chain,minimum size=1cm] (sed) {$S_d$};
		\node [above = of se] {$\mu$};
		\node [above = of se2] {$\mu$};
		\node [above = of sed] {$\mu$};

		% the arrows and labels
		\draw[->] (se) edge node[above] {} (wa5.west);
		\draw[->] (se2) edge node[above] {} (wadot.west);
		\draw[->] (wadot) edge node[above] {} (wad.west);
		\draw[->] (sed.east) -- +(30pt,0);
		
		\draw[<-] (wa.west) -- +(-20pt,0) node[left] {$\lambda_1$};

	\end{tikzpicture}
\caption{A network of tandem queues.}
\label{fig:tandemHD}
\end{figure}
\label{sec:many} 
Figure \ref{fig:tandemHD} pictures a network model with $d$ job classes and $d$ single-server stations, extending the tandem queues example presented in Section \ref{sec:tandem}. Class 1 jobs arrive from outside the system according to a Poisson process with rate $\lambda$, and they are served at station 1 on a FIFO basis. After completing service at the $i^{th}$ station, class $i$ jobs become class $i+1$ jobs and proceed to station $i+1$, where they are again served on a FIFO basis, for $i=1, \ldots, d-1$. Class $d$ jobs leave the system after completing service. Service times at each station are i.i.d. exponential random variables with mean $m < \lambda^{-1}$, and any service can be interrupted at any time and resumed later without any efficiency loss.

As in Section \ref{sec:tandem}, we assume that any server can be idled at any time, and there is no other mode of control available. This leads to a $d$-dimensional EWF of the form (\ref{problem:state:w})-(\ref{min:objective:new}) with the parameter values specified below. Adopting the sequence index $n = 400$ gives the drift vector
\begin{eqnarray*}
    \xi = (\sqrt{n}(\lambda - m^{-1}), 0, \ldots, 0) = (-1,0,\ldots,0).
\end{eqnarray*}
The covariance matrix is
\begin{equation*}
A=\left[ 
\begin{array}{ccccc}
2 & -1 &  &  &  \\ 
-1 & 2 & -1 &  &  \\ 
& -1 & \ddots & \ddots &  \\ 
&  & \ddots & \ddots & -1 \\ 
&  &  & -1 & 2%
\end{array}%
\right] ,
\end{equation*}%
and the control matrix is
\begin{equation*}
G=\left[ 
\begin{array}{ccccc}
1 &  &  &  &  \\ 
-1 & 1 &  &  &  \\ 
& -1 & \ddots &  &  \\ 
&  & \ddots & \ddots &  \\ 
&  &  & -1 & 1%
\end{array}%
\right] .
\end{equation*}
As in Section \ref{sec:tandem}, we assume the holding cost function is linear, that is,
\begin{align*}
    h(w) = h_1w_1 + \cdots + h_d w_d, \ w \in \mathbb{R}_+^d.
\end{align*}
The numerical values for holding cost rates $h_1,\ldots,h_d$ will be specified below. The control cost vector is $c=0$ (that is, there is no direct cost to idle any server), and the interest rate for discounting in the original model is $r = 0.01$, so the interest rate for our EWF is
\begin{align*}
    \gamma = n r = 4.
\end{align*}
Using our computational method, one can solve problems up to dimension $d \geq 30$, but solving the exact MDP formulation for such problems is not feasible computationally. Therefore, we look for effective benchmark policies, but tuning benchmarks is often computationally demanding as well. Thus, to ease the search for an effective benchmark, we assume $d$ is even and make the following assumption on the holding cost rate parameters:
\begin{equation}
h_2 > h_1 > h_4 > h_3>\cdots > h_{2k}> h_{2k - 1}> \cdots > h_{d} > h_{d-1}.
\label{eq:tandem:holding}
\end{equation}
It is intuitively clear that if $h_j>h_l$ for all $l>j$ and $j=1,\ldots,d-1$, then the control for station $j$ should be minimal, that is, server $j$ should keep working unless its buffer is empty. Therefore, we restrict attention to benchmark policies under which the control for even numbered stations is minimal, that is, servers $2,4,\ldots,d$ are idled only when their buffers are empty. We further restrict attention to benchmark policies under which the control for an odd-numbered station depends only on its own workload and the workload of the following station. To be more specific, for an odd numbered station station $i$, its server idles at time $t \geq 0$ if either its buffer is empty or the following holds:
\begin{eqnarray*}
\beta _{i}W_{i}(t) + 1\leq \beta _{i+1}W_{i+1}(t).
\end{eqnarray*}%
We refer to this benchmark policy as a linear boundary policy (LBP) and note that it is similar to the benchmark policy considered in Section 6.4 of \citet{ata2023drift}. 

\begin{figure}[!ht]
\subfigure[Dimension 1]{
		\includegraphics[width=2.in]{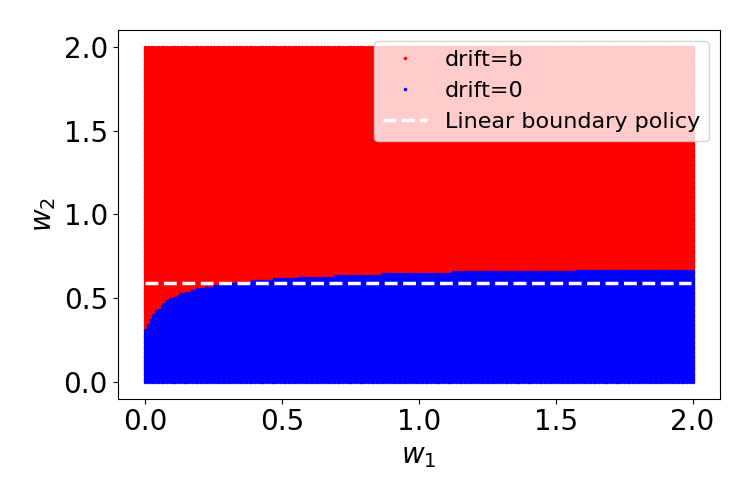}
  \label{fig:6d:tandem:1}} 
  \subfigure[Dimension 3]{		\includegraphics[width=2.in]{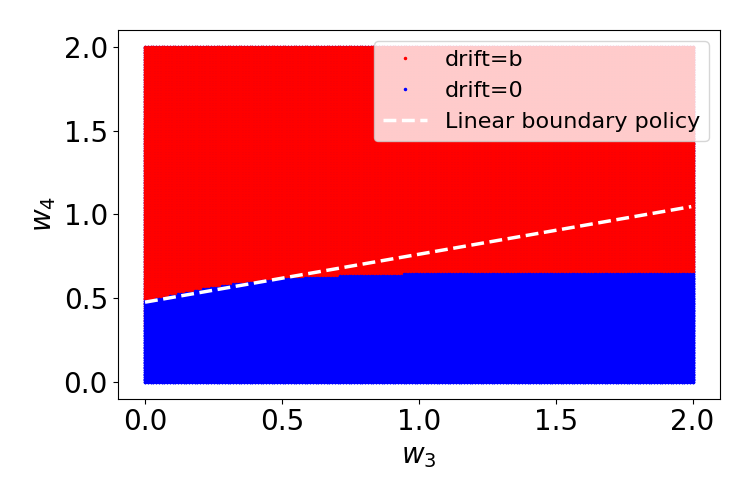}\label{fig:6d:tandem:2}}
    \subfigure[Dimension 5]{
  \includegraphics[width=2.in]{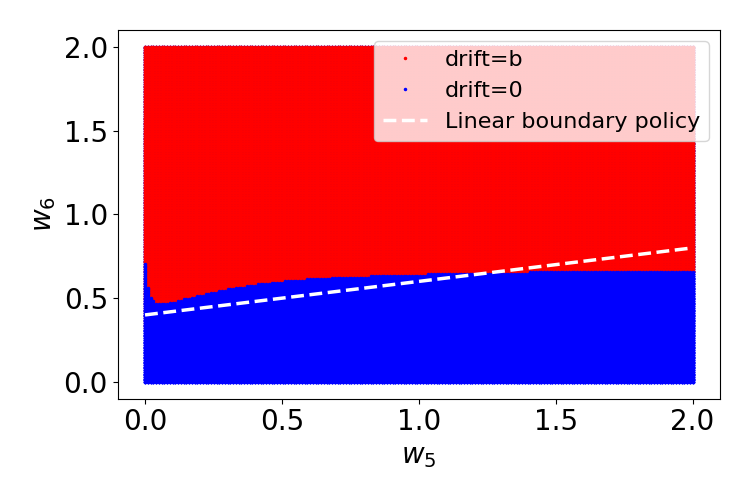}\label{fig:6d:tandem:3}}
\caption{Optimal linear boundary policy and our proposed policy for six queues in series.}
\label{fig:6d:tandem}
\end{figure}

For our numerical example, we let $d=6$ and
\begin{eqnarray*}
    h = (3, 3.9, 2, 2.9, 1, 1.9).
\end{eqnarray*}
Applying our computational method to this singular control problem, we set
the upper bound on the feasible drift rates at $b=20$. The resulting optimal policy is illustrated in Figure \ref{fig:6d:tandem} and compared with the best linear boundary policy. More precisely, the comparison is for states  $(w_1,w_2,1,1,1,1),(1,1,w_3,w_4,1,1)$ and $(1,1,1,1,w_5,w_6)$ in panels (a), (b) and (c) of Figure \ref{fig:6d:tandem}, respectively. The policy derived from our method closely aligns with the best linear boundary policy. It takes about ten hours of computing to find the optimal policy in our method using a 30-CPU core machine.

Table \ref{tab:simulation:tandem:6d} presents the simulated performance (that is, the average present value of holding costs incurred over the infinite planning horizon) for both policies implemented in our original queueing network, including standard errors for comparison; see Appendix \ref{appendix:pre-limit:policies} for a precise description of our proposed policy. Additionally, Table \ref{tab:simulation:tandem:6d} 
shows the performance of the policy that never idles any servers unless its buffer is empty. These simulations show that the policy we propose, based on the singular control approximation, performs about as well as the best linear boundary policy.
\begin{table}[!htb]
\centering
\begin{tabular}{ccc}
\toprule Never idle & Linear boundary policy & Proposed policy \\ 
\midrule 7011 $\pm$ 2.8 & 6924 $\pm$ 2.7 & 6933 $\pm$ 2.7 \\ 
\bottomrule
\end{tabular}%
\caption{Simulation performance for six queues in series.}
\label{tab:simulation:tandem:6d}
\end{table}

\section{Concluding remarks}
\label{sec:conclude}
Expanding on a statement made in Section \ref{sec:intro} of this paper, one may say that there are three steps involved in formulating a queueing network control policy based on a heavy traffic singular control approximation: (a) formulate the approximating singular control problem and derive its equivalent workload formulation (EWF), specifying the data of the EWF in terms of original system data; (b) solve the EWF numerically; and (c) translate the EWF solution into an implementable policy for the original discrete-flow model. The first of those steps has been the subject of research over the past 35 years, starting with \citet{harrison1988brownian} and including \citet{harrison1997dynamic}. The second step is the subject of this paper. For three of the four queueing network examples treated in this paper, the third step (or at least one version of the third step) is spelled out in Appendix \ref{appendix:pre-limit:policies} below, but we have not presented a general resolution of the translation problem. 

\citet{harrison1996bigstep} proposes a general framework for the translation task using discrete-review policies. To be specific, the proposed policies review the system status at discrete points in time and solve a linear program to make control decisions for the original queueing network. In addition to the first-order problem data, the linear program also uses the gradient $\nabla V$ of the optimal value function for the approximating EWF. \citet{harrison1996bigstep} does not attempt a proof of asymptotic optimality of the proposed policy, but in a follow up paper (\citet{harrison1998discrete-review}), the author considers a simple parallel-server queueing network and shows that the framework proposed in \citet{harrison1996bigstep} yields a policy for the original queueing network that is asymptotically optimal in the heavy traffic regime; also see \citet{harrison-lopez1999}. For the analysis of discrete-review policies, the state-of-the-art is \citet{ata-kumar2005}. The authors prove the asymptotic optimality of a discrete-review policy for a general queueing network under the so called complete resource pooling assumption.

A common feature of the formulations in these papers is that their approximating EWF is one-dimensional, and it admits a pathwise optimal solution. In particular, one can drop the gradient $\nabla V$ of its value function from the aforementioned linear program that is the crux of the framework proposed in \citet{harrison1996bigstep}. On the other hand, the solution approach proposed in this paper broadens the class of singular control problems one can solve significantly. As such, it paves the way for establishing the asymptotic optimality of the discrete-review policies in full generality as proposed in \citet{harrison1996bigstep}, providing a general resolution of the translation problem that we intend to revisit in future work.

All of the queueing network examples treated in this paper assume independent Poisson arrival streams and exponential service time distributions, but a major virtue of heavy traffic diffusion approximations is that they are distribution free. That is, the approximating diffusion model (in our setting, the approximating singular control problem) depends on the distributions of the original queueing network model only through their first two moments. Our primary reason for using Poisson arrivals and exponential service time distributions in our examples is to allow exact MDP formulations of the associated queueing control problems, which can then be solved numerically for comparison against our diffusion approximations.

The following two paragraphs describe extensions of the problem formulation
propounded in Section \ref{sec:general} of this paper, both of which are of
interest for queueing applications. Each of them is straight-forward in
principle but would require substantial effort to modify existing code.

\textbf{Weaker assumptions on the structure of the control matrix $G$.} To
ensure the existence of feasible controls, we assumed in Section \ref%
{sec:general} that the first $d$ columns of the control matrix $G$
constitute a Minkowski matrix. That is a relatively strong assumption
carried over from \citep{ata2023drift}, whose computational method we use to
solve our approximating drift control problem. A weaker assumption is that
those first $d$ columns constitute a \textit{completely-S} matrix, which %
\citet{taylor1993existence} showed is a sufficient condition for existence
of feasible controls in our setting. It seems very likely that our
computational method for drift control, and hence our problem formulation in
Section \ref{sec:general}, can be generalized to the completely-S case, and
that extension is definitely of interest for queueing applications.

\textbf{More general state space for the controlled process $W$.} To
accommodate closed network models, or networks whose total population is
bounded, as well as finite buffers and other queueing model
features, one would like to to allow a state space for $W$ that is a general 
$d$-dimensional polytope. This extension also appears to be routine in
principle, subject to the following two restrictions: first, the polytope is 
\textit{simple} in the sense that no more than $d$ boundary hyperplanes
intersect at any of its vertices; and second, at every point on the boundary
of the polytope, there exists a convex combination of available directions
of control that points into the interior from that point. The latter
requirement can be re-expressed in terms of completely-S matrices.

Lastly, we focused only on the discounted cost criterion for brevity, but our approach can be used to solve the singular control problem in the ergodic case too, using the corresponding algorithm of \citet{ata2023drift} and making minor changes to our overall approach and the code.

\bibliographystyle{plainnat}
\bibliography{mybib}
%\addcontentsline{toc}{section}{\refname}
\begin{appendix}

    \section{Implementation details}
    \subsection{Our policies}\label{app:our-policies}
							  In Section \ref{sec:drift}  of this paper, we have explained a method for approximating a singular control problem by a related problem of drift control. We begin this appendix by summarizing the computational method developed in our previous paper \citep{ata2023drift} for solving such drift control problems. Here we use the notation established in Sections \ref{sec:general} and \ref{sec:drift} of the current paper, which differs slightly from the notation of \citet{ata2023drift}. In particular, we denote by $W=\{W(t),t \geq 0\}$ the \textit{state process} to be controlled, with generic state $w\in \mathbb{R}_+^d$. 

    For either of the two drift control 
    formulations presented in Section 
    \ref{sec:drift}, a first step is to choose a constant drift rate $\tilde{\theta} 
    \in[0,b]^p$ that serves as a 
    \textit{nominal control policy}. That is, we use the term “nominal control policy” to 
    mean the policy that employs drift vector  $\tilde{\theta}$ in every state. We denote by 
    $\tilde{W}=\{\tilde{W}(t),t\geq 0\}$ the state 
    process under that nominal control policy.
    referring to $\tilde{W}$ as the \textit{reference 
    process} for our computational method. Also, the underlying Brownian motion $X$ that was 
    defined in Section \ref{sec:general} via 
    (\ref{problem:state:X}) will here be represented as 
    $X(t)=B(t)+\xi t$ for $t \geq 0$, where $B$ is a $d$-dimensional Brownian motion with covariance matrix $A$, zero drift and $B(0)=0$. Thus our reference process 
    $\tilde{W}$ has the representation $$\tilde{W}(t)=B(t)+\xi t+G \tilde{\theta} t+R \tilde{Y}(t).$$

    Let us consider the drift control formulation \textit{with} exogenous reflection at the boundary, which includes a corresponding vector $\pi$ of boundary penalty rates. Building on earlier work by \cite{han2018solving}, one can extend slightly the argument presented in Section 4 of \citet{ata2023drift} to arrive at the stochastic differential equation (\ref{app:eqn:sde}) below, in which $V(\cdot)$ is the optimal value function (initially unknown) for our discounted drift control problem, and the time horizon $T>0$ can be chosen arbitrarily. In the end, the search for an optimal drift control policy is reduced to solving the following reference SDE:
\begin{align}
    &e^{-\gamma T} V\left(\tilde{W}(T)\right)-V\left(\tilde{W}(0)\right) \label{app:eqn:sde}\\
    &=\int_0^T e^{-\gamma t} \nabla V(\tilde{W}(t)) dB(t) - 
    \int_0^T e^{-\gamma t} \{h(\tilde{W}(t))+g(\nabla V(\tilde{W}(t)))\}dt - \int_0^Te^{-\gamma t} \pi \cdot d  Y(t), \notag
\end{align}
    where $g(u)=\min_{\theta\in [0,b]^p}\{(c+G^\top u) \cdot \theta\}-(G^\top u)\cdot \tilde{\theta}$ for $u \in \mathbb{R}^d_+$. Our goal is to find a sufficiently regular function $V:\mathbb{R}^d_+ \rightarrow \mathbb{R}$ such that the identity (\ref{app:eqn:sde}) holds for almost every sample path $\{\tilde{W}(t),0\leq t\leq T\}.$ For the drift control formulation \textit{without} exogenous reflection at the boundary, the computational procedure is exactly the same, except that the vector $\pi$ appearing in the final term on the right side of (\ref{app:eqn:sde}) consists of the first $d$ components of the control cost rate vector $c$ (see Section \ref{sec:drift}).
    
    Our computational method proceeds as follows. First, for the chosen time horizon $T > 0$, we independently sample $N$ discretized paths of the Brownian motion $\{B(t), 0 \leq t \leq T\}$, 
    using a discretization step size $\Delta t > 0$. (Hereafter $N$ will be referred to as a "batch size.") We then construct the corresponding discretized paths of the reference process
    $\{\tilde{W}(t),0\leq t \leq T\}$. 
    Also, we represent the functions $(V,\nabla V)$ by neural networks $(V_{w_1},G_{w_2})$ that are  parameterized by weight vectors $\{w_1,w_2\}$. Let $\Delta_n(w_1,w_2)$ be the difference between the left and right sides of the reference SDE when (a) $B$ and $\tilde{W}$ are replaced by their $n$th discretized samples, (b) $V$ and $\nabla V$ are replaced by $V_{w_1}$ and $G_{w_2}$, and (c) the integrals on the right-hand side are replaced by the obvious approximating sums. Finally, we use a standard optimizer to choose values for $w_1$ and $w_2$ that minimize the sum of $\Delta_n^2(w_1,w_2)$ over $n=1,\cdots,N$.
    
     The problem-specific hyper-parameters that were used in training the neural networks for our examples are listed in Table \ref{tab:hyper-thin}, and the hyper-parameters that are common to all our examples are specified immediately below.
    
    \textbf{Common hyperparameters.} We chose the batch size $N=256$, the time horizon $T=0.1$, and the discretization step size $ \Delta t =0.1/64$.

\textbf{Optimizer.} We used the Adam optimizer \citep{kingma2014adam}.

\textbf{Activation function.} We use the 'elu' action function \citep{rasamoelina2020review}.

We modified the algorithm described in \citet{ata2023drift} to speed up the neural network training. These modifications exploit the special structure of our examples and are described below.
\begin{table}[!htbp]
    %\captionsetup{skip=0pt}
		{\small{ 
				\centering
				\begin{tabular}{lllllll}
					\toprule
					{Hyperparameters} & Parallel & Tandem & Criss-cross & Three-station & Many queues \\
					
					\midrule
     Section & \ref{sec:known} & \ref{sec:tandem} & \ref{sec:criss-cross}  & \ref{sec:BIGSTEP example} & \ref{sec:many} \\
     Dimension & 30 & 2 & 2 & 2 & 6  \\
     b   & 10 & 20 & 20 & 200 & 20 \\
     Reference policy $\tilde{\theta}$ & -1.0 & 0.0 & [0.0,0.5] & 0.0 & $[0.5,\ldots,0.5]$ \\
					\#Iterations & 6000 & 6000 & 6000 & 6000 & 9000\\
					\#Epoches & 135    & 34  & 34       & 40    & 52\\
					\multirow{5}[0]{*}{Learning rate  scheme} & {\footnotesize{5e-4 (0,3e3)}} & {\footnotesize{5e-4 (0,3e3)}} & {\footnotesize{5e-4 (0,3e3)}} & {\footnotesize{5e-4 (0,1.9e4)}} & {\footnotesize{5e-4 (0,9.5e3)}}\\
					& {\footnotesize{3e-4 (9.5e3,2.2e4) }} &  {\footnotesize{3e-4 (3e3,9e3)}} & {\footnotesize{3e-4 (3e3,9e3) }} & {\footnotesize{3e-4 (1.9e4,4.4e4) } } & {\footnotesize{3e-4 (3e3,9e3)}} \\
					& {\footnotesize{1e-4 (2.2e4,$\infty$)}}  & {\footnotesize{1e-4 (9e3,$\infty$)}} & {\footnotesize{1e-4 (9e3,$\infty$)}} & {\footnotesize{1e-4 (4.4e4,7e4) }} & {\footnotesize{1e-4 (9e3,1e5)}}\\
     & & & & 3e-5 (7e4,$\infty$) & {\footnotesize{5e-5 (1e5,1.5e5)}} \\
     & & & & & {\footnotesize{2e-5 (1.5e5,$\infty$)}} \\
					\#Hidden layers & 3 & 3 & 3 & 3 & 3  \\
					\#Neurons in each  layer & 300 & 50 & 50 & 100 & 100 \\
					\bottomrule%
				\end{tabular}%
		}}
  \caption{Hyperparameters used in the numerical examples.}
				\label{tab:hyper-thin}%
	\end{table}
 
\textbf{Decay loss.}  
For the parallel queue example treated in Section \ref{sec:known}, we implement the same decay loss approach that was used in Section E.1 of \citet{ata2023drift}. Here we let
 $$\tilde{b} =\left(7 - \frac{\textrm{iteration}}{4800}\right)^+.$$

    \textbf{Increasing upper bound $b$.} For the criss-cross network (Section \ref{sec:criss-cross}), the three-station example (Section \ref{sec:BIGSTEP example}), and the many queues in series example (Section \ref{sec:many}), we gradually increase the upper bound $b$ during the neural network training. We start from a small value and gradually increase it to the target level. To be specific, we let

\[
\hat{b}=\left\{ 
\begin{array}{ll}
\min \left( 20,4+\frac{\textrm{iteration}}{80}\right)  & \text{for the criss-cross network,} \\ 
\min \left( 200,40+\frac{\textrm{iteration}}{800}\right)  & \text{for the three-station network,} \\ 
\min \left( 20,4+\frac{\textrm{iteration}}{14400}\right)  & \text{for the many queues in series example.}%
\end{array}%
\right. 
\]
    
\textbf{Shape constraints.}  Let $V:\mathbb{R}^d_+ \rightarrow \mathbb{R}$ denote the optimal value function for the singular control problem (\ref{problem:state:w})-(\ref{min:objective:new}) introduced in Section \ref{sec:general}. That is, let $V(w)$ denote the minimum achievable expected present value of costs incurred over the infinite planning horizon, given that the initial state vector is $W(0)=w$. The computational method that we proposed in \citet{ata2023drift} provides not only an approximation for $V$ but also an approximation $G :\mathbb{R}^d_+ \rightarrow \mathbb{R}^d$ for the gradient of $V$. In both cases, the approximation comes in the form of a neural network. In order to emphasize this, we will write 
$G^{\theta}(w) = (G^{\theta}_1(w), \ldots,G^{\theta}_d(w))$ where $\theta$ denotes the neural network parameters.

For the examples considered in Sections \ref{sec:tandem}- \ref{Sec:Many queues}, one can infer certain properties of the value function and its gradient, which we use to speed up the training process. Next, we describe these properties, which we refer to as the shape constraints, for each example.
\begin{itemize}
    \item \textbf{Criss-cross network example.} In all four cases considered in Section \ref{section:criss-cross}, one observes the following from Table \ref{tab:criss-cross:cost function}: i) When $2w_1 > w_2$, the holding cost function $h(w_1,w_2)$ is increasing in $w_1$, and ii) when $2w_1 \leq w_2$, it is increasing in $w_2$. Intuitively, these suggest that
\begin{align*}
\frac{\partial V(w)}{\partial  w_1} >0  & \ \text{ when } \ 2w_1>w_2, \\ 
\frac{\partial V(w)}{\partial  w_2} > 0 & \ \text{ when }\  2w_1\leq w_2.
\end{align*}
Thus, we seek a neural network approximation that satisfies these "shape constraints" or monotonicity requirements. That is, we seek a neural network approximation that satisfies
\begin{align*}
G_1^{\theta}(w) >0  & \ \text{ when } \ 2w_1>w_2, \\ 
G_2^{\theta}(w)> 0 & \ \text{ when }\  2w_1\leq w_2.
\end{align*}
We pursue this goal by amending our loss function for the neural network training with the following penalty term:
\begin{eqnarray*}
  \left(   [ G_1^{\theta}(w) ]^- \, \mathbb{I}\{2w_1>w_2\}+[ G_2^{\theta}(w) ]^- \, \mathbb{I}\{2w_1\leq w_2\}   \right)^2,  
\end{eqnarray*}
where $\mathbb{I}\{ \cdot\}$ is the indicator function.

    \item \textbf{Three-station network example.} Recall that the holding cost funcion $h(w) = w_1 + w_2$ for this example. Because $h(w)$ is monotone, one expects intuitively that $\nabla V(w) \geq 0$. Thus, we look for a neural network approximation that has $G^{\theta}(w) \geq 0$ for $w \in S$. To do so, we add the following penalty term to the loss function used for neural network training: 
\begin{eqnarray*}
    \left([ G_1^{\theta}(w) ]^- \right)^2 + \left([ G_2^{\theta}(w) ]^- \right)^2.
\end{eqnarray*}

    \item \textbf{Tandem queue and many queues in series examples.} As described in Appendix \ref{appendix:pre-limit:policies}, server $i$ ($i=1, \ldots, d-1$) idles if
    \begin{eqnarray}
        \frac{\partial V(w)}{\partial w_i} - \frac{\partial V(w)}{\partial w_{i+1}} < 0. \label{eqn:tandem:shape:constraint}
    \end{eqnarray}
Intuitively, idling server $i$ becomes less appealing if its workload increases. Thus, one expects the left-hand side of (\ref{eqn:tandem:shape:constraint}) to be increasing in $w_i$. That is,
\begin{eqnarray*}
    \frac{\partial}{\partial w_i} \left[ \frac{\partial V(w)}{\partial w_i} - \frac{\partial V(w)}{\partial w_{i+1}}   \right] >0.
\end{eqnarray*}
Therefore, we seek a neural-network approximation that satisfies
\begin{eqnarray*}
    \frac{\partial}{\partial w_i} [G_i^{\theta}(w) - G_{i+1}^{\theta}(w)] > 0, \  i=1,\ldots,d-1.
\end{eqnarray*}
To do so, we add the following penalty term to the loss function
\begin{eqnarray*}
    \sum_{i=1}^{d-1} \left[ \left( \frac{\partial}{\partial w_i} [G_i^{\theta}(w) - G_{i+1}^{\theta}(w)] \right)^- \right]^2.
\end{eqnarray*}
 
\end{itemize}
  \subsection{MDP solutions}
  For the tandem queues example (Section \ref{sec:tandem}) and the criss-cross network example (Section \ref{sec:criss-cross}), we solve the MDP associated with the original queueing network models via value iteration. For each example, we truncate the state space by imposing an upper bound for each buffer. To be specific, we let the upper bound be 1000 for the tandem queues example and 300 for the criss-cross example. In both cases, we stop the value iteration when the iteration error is less than $\epsilon =0.1$.
  
    \subsection{A benchmark for the three-station example via \citet{kushner1991numerical}}
In addition to solving the singular control problem for the three-station network example (Section \ref{sec:BIGSTEP example}) using our method, we also solve it by adopting the method of \citet{kushner1991numerical}, which builds on \citet{kushner1990numerical}. This provides a benchmark for comparison with our method. Note that our covariance matrix does not satisfy the assumption (5.5) in \citet{kushner1990numerical}. Therefore, we used different grid sizes in different dimensions that satisfy (5.9) in \citet{kushner1990numerical}. Specifically, the discretization grid (using the notation in \citet{kushner1990numerical}) is defined with $h_1=0.1$ in the first dimension and $h_2=0.108$ in the second dimension; and the state space is truncated at $400 h_1 = 40$ in the first dimension, and it is truncated at $400 h_2 =43.2$ in the second dimension.
    
    \subsection{Linear boundary policies}
    As mentioned in Section \ref{sec:many}, we use the linear boundary policy (LBP) as the benchmark for the many queues in series example. To design an effective LBP, we perform a grid search over its parameters $\beta_1, \ldots, \beta_6$. In particular, we consider the following values:
    \begin{align*}
         &\beta_i \in \{0, 0.1, \ldots, 1.0 \}  \text{ \ for } i=1,3,5, \\
         &\beta_i \in \{0.4, 0.5, \ldots, 3.6 \}   \text{ for } i=2,4,6.
    \end{align*}
  The range of values considered are informed by a preliminary search over a broader range of values. A brute-force search over all possible parameter combinations involves evaluating $(11\times33)^3$ scenarios, which is prohibitive computationally. Therefore, we proceed with the following heuristic approach:
    \begin{enumerate}
        \item We search $\beta_1$ and $\beta_2$ assuming servers 3 and 5 are non-idling.
        \item Given the best performing $\beta_1$ and $\beta_2$ obtained in the previous step, we then search for $\beta_3$ and $\beta_4$ assuming server 5 is non-idling.
        \item Given the best performing $\beta_1, \ldots,\beta_4$ obtained in the previous two steps, we then search for $\beta_5$ and $\beta_6$.
    \end{enumerate}
    This heuristic method reduces the number of cases to $3\times11\times33$. For each case, we simulate 300000 independent paths. The resulting parameters are as follows:
    $$ (\beta_1, \ldots, \beta_6)= (0.0,1.7,0.6,2.1,0.5,2.5).$$

\label{sec:app:lbp}
    \section{Details of the simulation studies for policy comparisons}
    We performed 400,000 independent replications for all examples considered except for the parallel-server queueing network example (Section \ref{sec:known}). For that example, we performed 100,000 replications.
     
    \textbf{Further details for the tandem queues, the criss-cross network and the many queues in series examples.} For these examples, we simulate the original queueing network models, which avoids discretization errors. Also, we choose the time horizon for the simulation sufficiently long to approximate the infinite horizon discounted performance. To be specific, we make sure each simulated path is run longer than 1400 time units. Because the interest rate $r = 0.01$, the discount factor at the termination is less than or equal to $\exp{-1400 \, r} \approx 8 \, e^{-7}$, indicating a negligible error due to truncation of the time horizon.

    \textbf{Further details for the parallel-server queueing network and the three-station queueing network examples.} For those examples, we simulate the approximating diffusion models (Section \ref{sec:general}) via discretizing the time  of the diffusion model; see Table \ref{tab:simulation:details}. Then the singular control is implemented using a unit-step control. To be specific, if the process crosses the state boundary, we repeatedly apply a unit-step control to push it back into state space until that happens.
      \begin{table}[!ht]
        \centering
        %\captionsetup{skip=0pt}
        \begin{tabular}{lll}
        \toprule
        &  Parallel-server queueing network  & Three-station network \\
        \midrule
        Time discretization & 3.125e-5 & 0.0015625 \\
        Step size for unit control & 0.01 & 0.1 (1st dim), 0.108 (2nd dim)\\
        \bottomrule%
        \end{tabular}
        \caption{The specifics of the discretization grids and control unit steps for the parallel queues and three-station queues}
        \label{tab:simulation:details}
    \end{table}

    \section{Scheduling policies for queueing network examples.}\label{appendix:pre-limit:policies}

Recall from Appendix \ref{app:our-policies} that $V:\mathbb{R}^d_+ \rightarrow \mathbb{R}$ denotes the optimal value function for the singular control problem (\ref{problem:state:w})-(\ref{min:objective:new}) introduced in Section \ref{sec:general}. That is, $V(w)$ denotes the minimum achievable expected present value of costs incurred over the infinite planning horizon, given that the initial state vector is $W(0)=w$. Also recall that the computational method that we proposed in \citet{ata2023drift} provides not only an approximation for $V$ but also an approximation $G :\mathbb{R}^d_+ \rightarrow \mathbb{R}^d$ for the gradient of $V$. (In both cases, the approximation comes in the form of a neural network.) It is this function $G(\cdot)=(G_1(\cdot),\cdots,G_d(\cdot))$ that we use in the algorithmic implementation of our proposed scheduling policies for queueing network examples in Sections \ref{sec:tandem}, \ref{sec:criss-cross} and \ref{Sec:Many queues}.

In each case, we observe the system state $Q(t)=q$ immediately after a state transition time $t \geq 0$ (that is, immediately after an arrival from outside the system or a service completion) and determine an action (that is, which servers to idle, if any, and which classes to serve at stations where there is a choice) that remains in place until the next state transition.

To spell out the mechanics of those implementations, we begin with the tandem queues example (Section \ref{sec:tandem}) and many queues in series (Section \ref{Sec:Many queues}), because their policy descriptions are simplest. In those examples, the last server keeps working unless its buffer is empty. Thus, we will focus on the upstream servers to describe our proposed policy. 

\textbf{Tandem queues example (Section \ref{sec:tandem}).} Denoting by $w$ a generic state of the workload process $W$, the phrase "$w$ lies in the region colored red in Figure \ref{fig:2d:tandem}" means precisely that $G_1(w) \leq G_2(w)$. In words, this means that an increase in $U_1$, which causes a displacement downward and to the right, effects a decrease in $V(\cdot)$, so increasing $U_1$ is favorable; gradient inequalities are interpreted similarly in the other two examples below. Thus, having observed that $Q(t)=q$ immediately after a state transition time $t \geq 0$, and recalling the definition $W(\cdot)=Q(n\cdot) / \sqrt{n}$, where $n$ is the sequence index or scaling parameter introduced earlier, our computational rule is the following: server 1 keeps working unless either $q_1 = 0$ or $ G_1(q / \sqrt{n}) \leq G_2(q/ \sqrt{n})$, in which case it idles.

\textbf{Many queues in series example (Section \ref{Sec:Many queues}).} Suppose that a $d$-dimensional queue length vector $Q(t)=q$ is observed immediately after a state transition time $t \geq 0$. For $i=1, \ldots, d-1$, server $i$ keeps working unless either $q_i = 0$ or 
$G_i(q/ \sqrt{n}) \leq G_{i+1}(q/ \sqrt{n})$, in which case it idles.

\textbf{Criss-cross network example (Section \ref{sec:criss-cross}).} We use the scheduling policy proposed by \citet{harrison1989scheduling} and \citet{martins1996heavy}. As discussed in Section \ref{section:criss-cross}, server 2 keeps working as long as its buffer isn't empty. Focusing attention on server 1, Algorithm \ref{algo:pre-limit:criss-cross} describes the scheduling policy. As a preliminary, recall that, given observations of the three-dimensional queue length process $Q(\cdot)$, we define $Z(\cdot) = n^{-1/2} Q(n\cdot)$, and define the associated two-dimensional (scaled) workload process $W(\cdot)=MZ(\cdot)$ .
\begin{algorithm}[!ht]
	\caption{Scheduling policy for the criss-cross network.}
	\label{algo:pre-limit:criss-cross}
	\begin{algorithmic}[1]
		\Require{Queue length vector $Q(t)=q$ observed immediately after a state transition time $t \geq 0$, the corresponding workload vector $w=n^{-1/2}Mq$, a safety stock parameter $s$, and the gradient function $G= (G_1,G_2)$ }
		\Ensure{A control action for server 1}
		\If{$q_1=0$ and $G_1(w)<0$ }
		\State idle server 1
		\Else
        \If{$q_3>s$ or $G_2(w)< 0$}
		\State give priority to class 1
		\Else 
		\State give priority to class 2
		\EndIf
		\EndIf
	\end{algorithmic}
\end{algorithm} 

\section{Analytical solution of a one-dimensional singular control problem}
\label{appendix:deri:1d}
We consider the one-dimensional singular control problem stated below. We assume $h> rc >0$. Otherwise, one can show that it is optimal not to exert any control.
\begin{eqnarray*}
&&\min_{U(\cdot)}E\left[ \int_{0}^{+\infty }e^{-rt}\left[ hW(t)+cdU(t)\right]  \right]  \\
&&\text{subject to} \\
&&W(t)=w+X(t)+Y(t)-U(t),\ t\geq 0, \\
&& Y, U \text{ are adapted to } X, \\ 
&&Y\text{ is continuous and non-decreasing with }Y(0)=0, \\
&&Y\text{ increases at times }t\geq 0\text{ when }W(t)=0, \\
&& U\text{ is right-continuous and non-decreasing with }U(0) \geq 0, \\
&& W(t) \geq 0, \ t \geq 0.
\end{eqnarray*}%
Using a smooth-pasting argument one can solve the corresponding HJB equation by considering the following problem; see for example, \citet{harrison1983instantaneous}: Find a threshold value $w^* > 0$ and 
$V: \mathbb{R}_+ \rightarrow \mathbb{R}$ that is twice continuously differentiable such that they jointly satisfy
\begin{eqnarray*}
\frac{a}{2}V^{\prime \prime }(w)+hw =rV(w),\text{ for } w \in \lbrack 0,w^{\ast }] \\
V^{\prime }(0) =0\text{ and }V^{\prime }(w)=c \text{ for }w \geq w^{\ast }.
\end{eqnarray*}
One can verify that the solution is of the following form:
\[
V^{\ast }(w)=\left\{ 
\begin{array}{c}
V_{1}(w) \\ 
V_{2}(w)%
\end{array}%
\right. 
\begin{array}{c}
\text{if }w<w^{\ast } \\ 
\text{if }w\geq w^{\ast }%
\end{array}%
, 
\]%
where 
\[
V_{1}(w)=\frac{h\sqrt{a}e^{-\frac{\sqrt{2}\sqrt{r}w}{\sqrt{a}}}}{\sqrt{2}%
r^{3/2}}+\frac{hw}{r}+C e^{\frac{\sqrt{2}\sqrt{r}w}{\sqrt{a}}}+C e^{-%
\frac{\sqrt{2}\sqrt{r}w}{\sqrt{a}}},\text{ and}
\]%
\[
V_{2}(w)=cw-cw^{\ast }+V_{1}(w^{\ast }),
\]%
for threshold $w^*$ and constant $C$ we specify next. Furthermore, the optimal control is that $U(t)$ increases only when $W(t)=w^{\ast }$
and confines the process in the interval $[0,w^{\ast }].$

To be specific, one needs to solve the following two equations to determine $w^*$ and $C$: 
\[
V_{1}^{\prime }(w^{\ast })=c\text{ and }V_{1}^{\prime \prime }(w^{\ast })=0.
\]%
For the parameter values we set forth in Section \ref{sec:known} (that is, $h=2,a=c=1,r=0.1$),
we have $w^{\ast } =0.722331$ and $C=-15.3786$.

\end{appendix}

\end{document}